%% file: paper_arxiv.tex
\documentclass[%
 aps,onecolumn,%
 amssymb, amsmath,
 preprint
]{revtex4-1}

\usepackage{docs}%
\usepackage{bm}%
\expandafter\ifx\csname package@font\endcsname\relax\else
 \expandafter\expandafter
 \expandafter\usepackage
 \expandafter\expandafter
 \expandafter{\csname package@font\endcsname}%
\fi
\usepackage[utf8]{inputenc}
\usepackage[T1]{fontenc} 
\usepackage[english]{babel}

\usepackage{graphicx}
\usepackage{siunitx}

\begin{document}

\title{Clustering of microscopic particles in constricted blood flow}%

\author{Christian B\"acher}
\author{Lukas Schrack}
\author{Stephan Gekle}
\affiliation{Biofluid Simulation and Modeling, University of Bayreuth, Universit\"atsstra{\ss}e 30, 95440 Bayreuth, Germany}%

\date{August 2016}%
%\revised{August 2010}%
\begin{abstract}
	A mixed suspension of red blood cells (RBCs) and microparticles flows through a cylindrical channel with a constriction mimicking a stenosed blood vessel.
	Our three-dimensional Lattice-Boltzmann simulations show that the RBCs are depleted right ahead and after the constriction.
	Although the RBC mean concentration (hematocrit) is 16.5\% or 23.7\%, their axial concentration profile is very similar to that of isolated tracer particles flowing along the central axis.
	Most importantly, however, we find that the stiff microparticles exhibit the opposite behavior.
	Arriving on a marginated position near the channel wall, they can pass through the constriction only if they find a suitable gap to dip into the dense plug of RBCs occupying the channel center.
	This leads to a prolonged dwell time and, as a consequence, to a pronounced increase in microparticle concentration right in front of the constriction.
	For biochemically active particles such as drug delivery agents or activated platelets this clustering may lead to physiologically relevant hot spots of biochemical activity.
\end{abstract}
\maketitle

\section{Introduction}

% short Intro
Blood mostly consists of highly deformable red blood cells (RBCs) occupying from 15 up to 45 volume percent \cite{Popel2005, Misbah2013, Gompper_2015, FreundRev2014}. 
When flowing through a channel or blood vessel, RBCs preferentially travel in the low-shear zone around the channel center.
Stiff particles such as white blood cells, platelets or synthetic microparticles are thereby forced to flow near the channel walls.
This well-known effect, called margination, is of physiological relevance, e.g., for platelets where a near-wall position allows them to quickly stop bleeding in case of vessel injury.
For synthetic particles such as drug delivery agents a near-wall position is essential to directly deliver their cargo to endothelial cells, thus minimizing unwanted distribution of pharmaceutical agents through the vascular system.

% former results
In 1980 margination was first observed experimentally \cite{SchmidSchoenbein1980}. 
Since then, a large number of studies investigated the phenomenon by means of experiments \cite{Eckstein1988,Jain2009,Charoenphol2010,Lee2013,Chen2013, Namdee_2013, Wang2013, Lee_2013_margination, Fitzgibbon_2015} 
or simulations \cite{Migliorini2002,Freund2007,Crowl2011, Kumar2011,Tokarev2011,  Zhao2011, Zhao2012,Kumar2012, Fedosov2012,Freund_2012, Reasor2013, Kumar2014,Fedosov_Gompper_2014,  Mueller2014,Vahidkhah_2014, Fitzgibbon_2015, Vahidkhah_2015, Rivera2015, Mueller2016, Gekle2016, Krueger2016, Mehrabadi_2016, Spann_2016}. 
All these works focus on spatially constant geometries like straight pipes, rectangular channels or plane-Couette systems.
However, more complex geometries such as bifurcations or constrictions (stenoses) are a common feature of the vascular system.
At the same time, investigations of marginated particles near such geometrical features are scarce.
For RBCs a recent detailed study in cylindrical microchannels with stenosis exists \cite{Vahidkhah_2016_stenosis}.
Ref. \citenum{Mountrakis2013} presented 2D simulations near aneurysms while ref.~\citenum{Zhao2008} investigated the recirculation zones after a sudden expansion.
Near a stenosis ref.~\citenum{Wang_2013_thrombus} presented local velocity fields while other studies treated the dynamics of platelets in 2D and plane-Couette geometries \cite{Skorczewski2013, Yazdani_2016} .

% What is done
Here, we use 3D Lattice-Boltzmann simulations to study margination of stiff spherical microparticles in a red blood cell suspension flowing through a cylindrical channel with a constriction mimicking the geometry of a stenosed blood vessel.
While the radial concentration profile is found to be qualitatively unaffected (RBCs in the center, microparticles near the wall) we find that the concentration profile along the axial direction varies significantly for both RBCs and microparticles.
Right in front of the constriction we observe a decrease in RBC concentration which can be almost quantitatively explained by the flow acceleration using passive point particles flowing along the central axis as the most simple model system.
Microparticles, on the other hand, exhibit the opposite behavior: their concentration in front of the constriction increases significantly.
This effect can be traced back to the near-wall trajectories on which microparticles travel in the straight channel section before arriving at the constriction and seems to be strongly reinforced by the cylindrical geometry since a recent study in planar geometries \cite{Yazdani_2016} did not report it.
In order to pass through the constriction, the particles need to leave this position and squeeze into the densely packed red blood cell layer.
This lowers the microparticles' passing rate leading to a pronounced peak in microparticle concentration ahead of the constriction.

\section*{Materials and Methods}
\input{materials_methods.tex}

\begin{figure*}
	\centering
	a)
	\begin{minipage}{0.25\textwidth}
		\includegraphics[width=\textwidth]{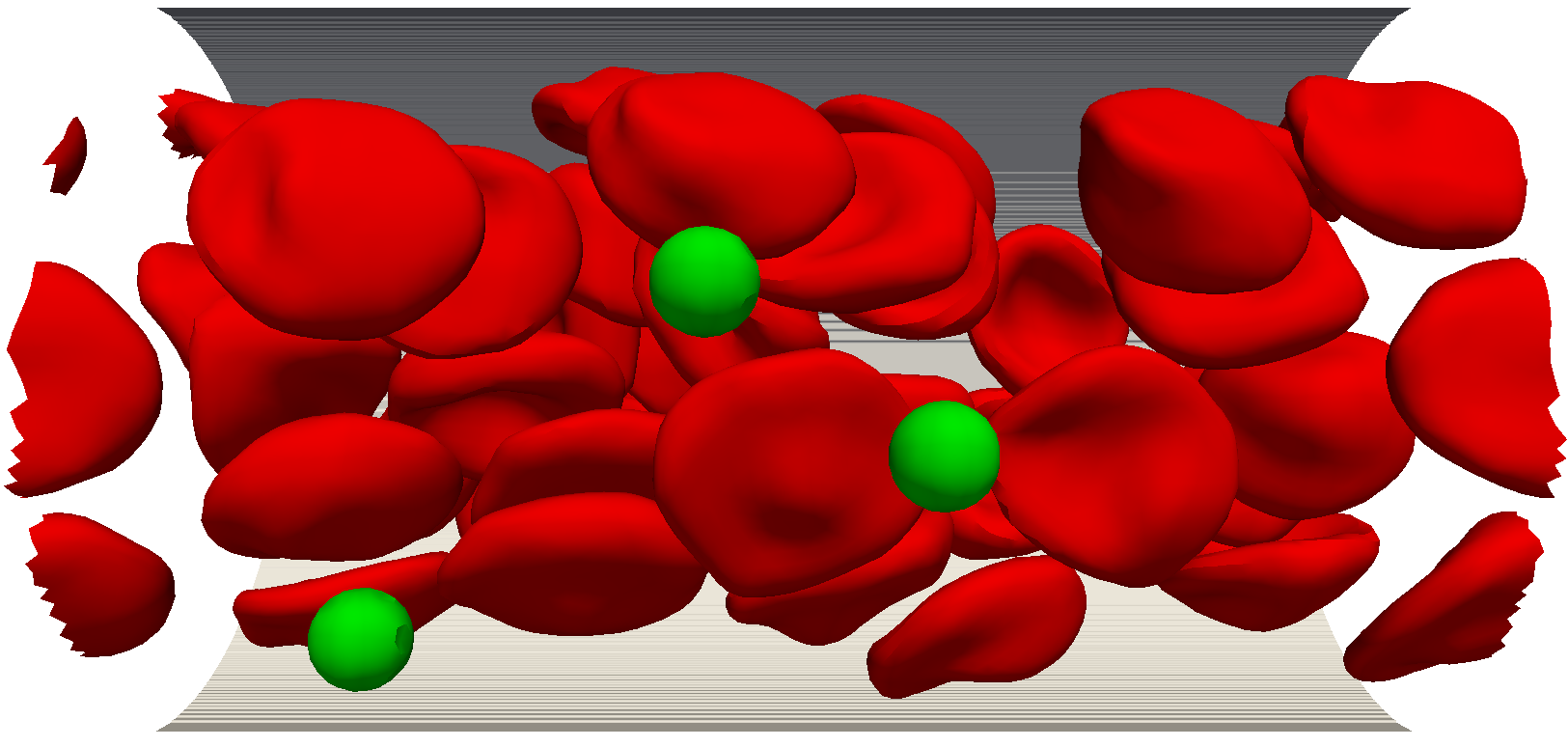}
	\end{minipage}
	b)
	\begin{minipage}{0.68\textwidth}
		\includegraphics[width=\textwidth]{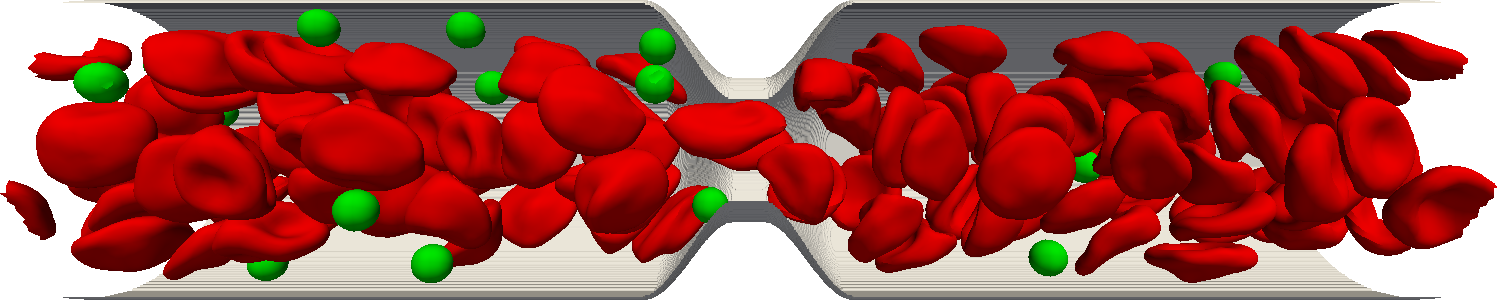}
	\end{minipage}
	
	c)
	\begin{minipage}{0.68\textwidth}
		\includegraphics[width=\textwidth]{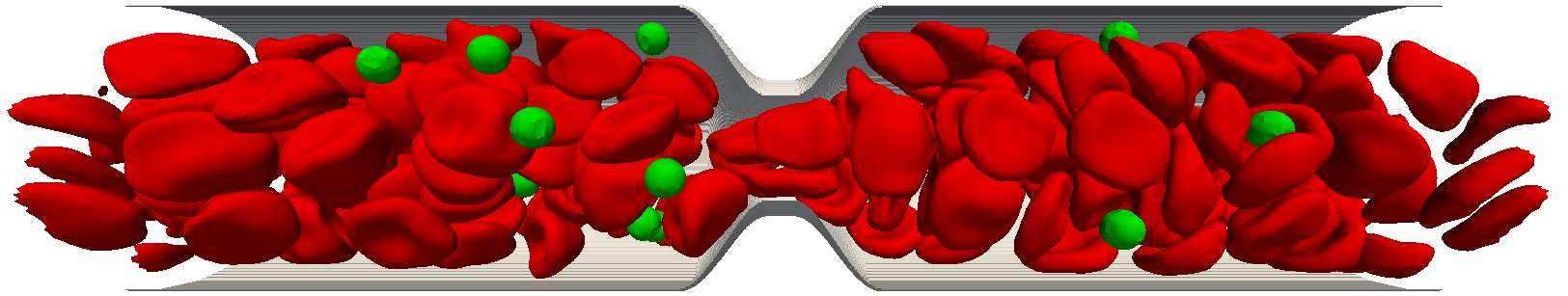}
	\end{minipage}
	\caption{Snapshot of the simulated system. (a) Straight cylindrical channel with 47 RBCs and 4 microparticles (Ht=16.5\%).
		(b) Constricted channel with 108 RBCs and 18 microparticles (Ht=16.5\%) and (c) with 151 RBCs and 18 microparticles (Ht=23.7\%). The ratio between the radii of the constricted and the straight cylinder is 1/2. A periodic boundary condition in flow direction is used. The flow is from left to right.}
	\label{FIG:system}
\end{figure*}

%%
%% Results
%%

\section*{Results and Discussion}

A straight channel as comparison and two examples of the investigated systems with stenosis are shown in figure~\ref{FIG:system}. 
In all cases the cylinder radius is $R_C = 13.4{\mu m}$ and in (b) and (c) the ratio between the radius of the constricted part and the main cylinder is 1/2. 
The length of the region with constant lower radius here is $2.6{\mu m}$ but will be varied below.
The length of the straight channel is $47{\mu m}$ and $117{\mu m}$ for the constricted channel.
We verify in the supplementary material that varying the channel length has no influence on the results.
The hematocrits are 16.5\% in (a) and (b) and 23.7\% in (c).
The mean RBC velocity is $1.7{\frac{mm}{s}}$ in the main channel in (a) and $1.3{\frac{mm}{s}}$ and $0.8{\frac{mm}{s}}$ inside the constriction in (b) and (c), respectively.

\subsection*{Center-of-mass distribution in comparison to straight cylinder}

\begin{figure*}
	\centering
	\includegraphics[width=\textwidth]{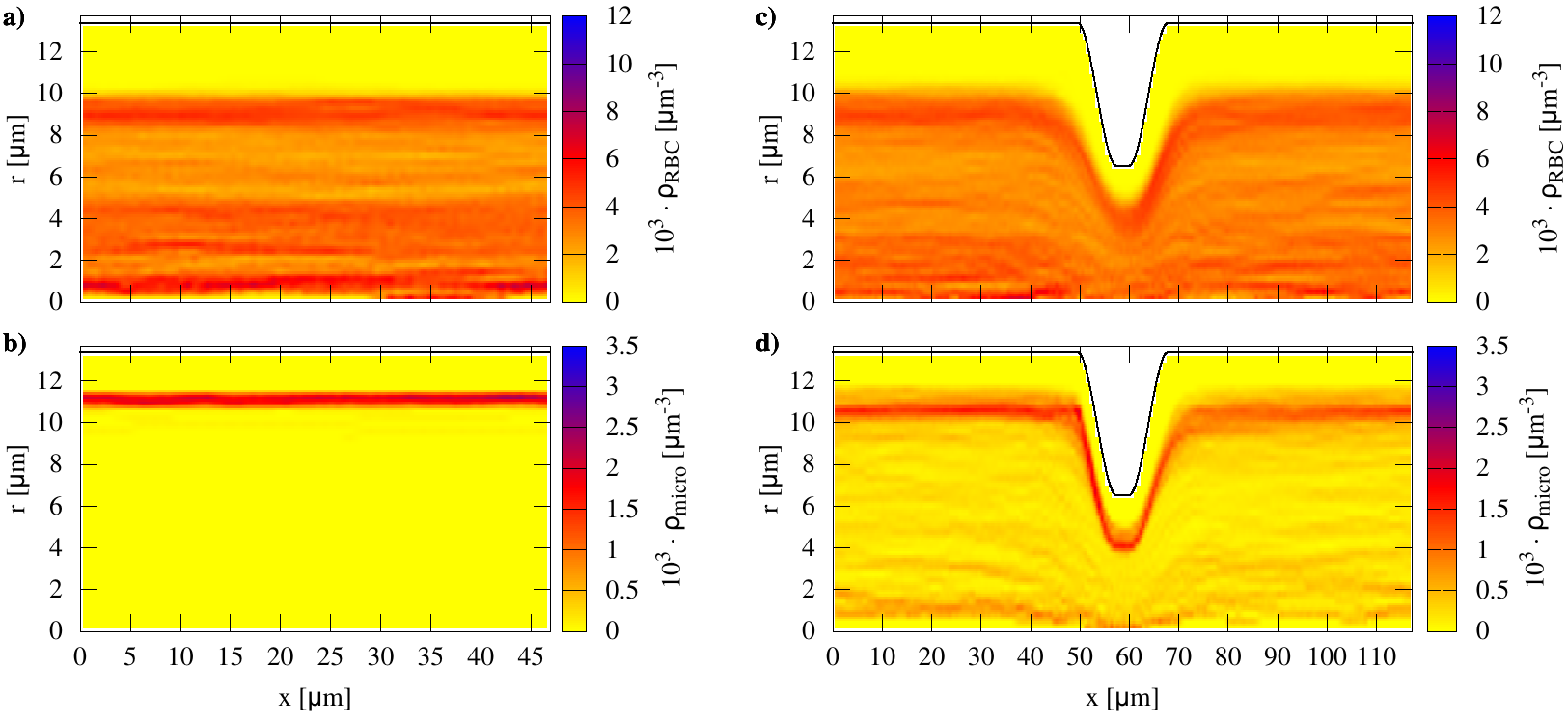}
	\caption{Concentration as a function of the radial and axial position plotted for the (a) RBCs and (b) microparticles in the straight cylinder and for the (c) RBCs and (d) microparticles in a cylinder with constriction. The channel geometry is illustrated by the black line.}
	\label{FIG:map}
\end{figure*}

We first extract two-dimensional concentration profiles in the $x$-$r$ plane, where $x$ is the coordinate along the channel and $r$ is the radial position in cylindrical coordinates. 
Local concentrations of RBCs and microparticles are calculated from their center-of-masses as described in the Methods section.
The result for the straight channel is shown in figure~\ref{FIG:map}~(a) for RBCs and in (b) for microparticles.
The RBC concentration is rather homogeneous and the RBCs are located around the channel center since their deformability causes the RBCs to migrate to the center \cite{Coupier_2008, Geislinger_2012, Grandchamp_2013}. 
Near the channel wall, the well-known cell-free layer with almost zero RBC concentration is visible \cite{Fedosov_2010_CFL, Freund_2011, Katanov_2015}.
As can be seen in figure~\ref{FIG:map}~(b) the microparticles are expelled into this cell-free layer leading to a sharply peaked concentration distribution of the marginated microparticles.
The origin of microparticle migration into the cell-free layer is commonly attributed to heterogeneous collisions with the RBCs \cite{Crowl2011,Tokarev2011,Kumar2012,Kumar2014,Rivera2015, Spann_2016, Yazdani_2016, Gekle2016}. 

The general picture looks similar for the constricted channel with Ht=16.5\% as shown in figure~\ref{FIG:map}~(c) and (d) (see supplementary material for Ht=23.7\%).
Besides a reduced cell-free layer inside the constriction, which agrees well with ref.~\cite{Vahidkhah_2016_stenosis}, the RBC concentration in the channel center is not significantly influenced by the constriction. 
The peak at large radii in the microparticle concentration is however more blurred compared to the straight channel, especially right behind the constriction. 
%A similar effect has been seen for planar channels \cite{Yazdani_2016} 
This can be explained since the microparticles require a certain time (or, equivalently, distance) in order to regain their fully marginated state after passing through the constriction.

\subsection*{Axial distribution in one-component suspensions}

\begin{figure*}
	\centering
	a)
	\begin{minipage}{0.3\textwidth}
		\includegraphics[width=\textwidth]{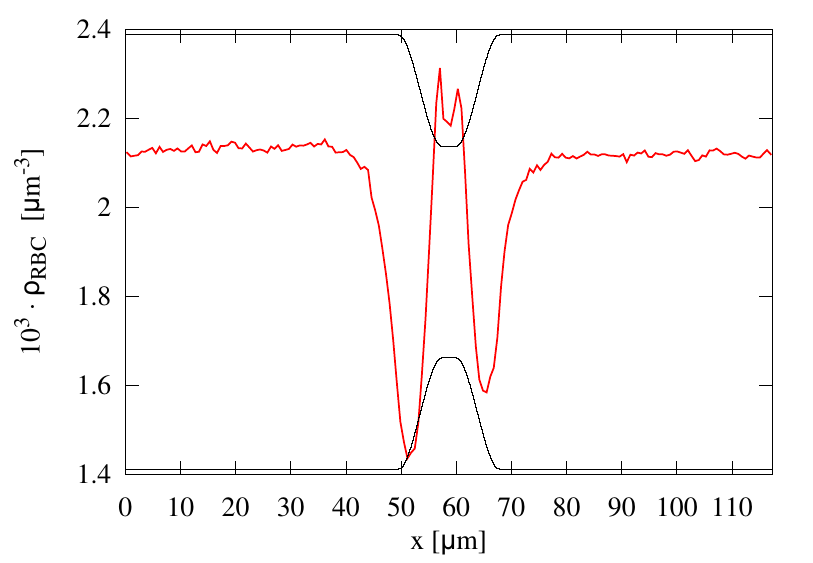} 	
	\end{minipage}
	b)
	\begin{minipage}{0.3\textwidth}
		\includegraphics[width=\textwidth]{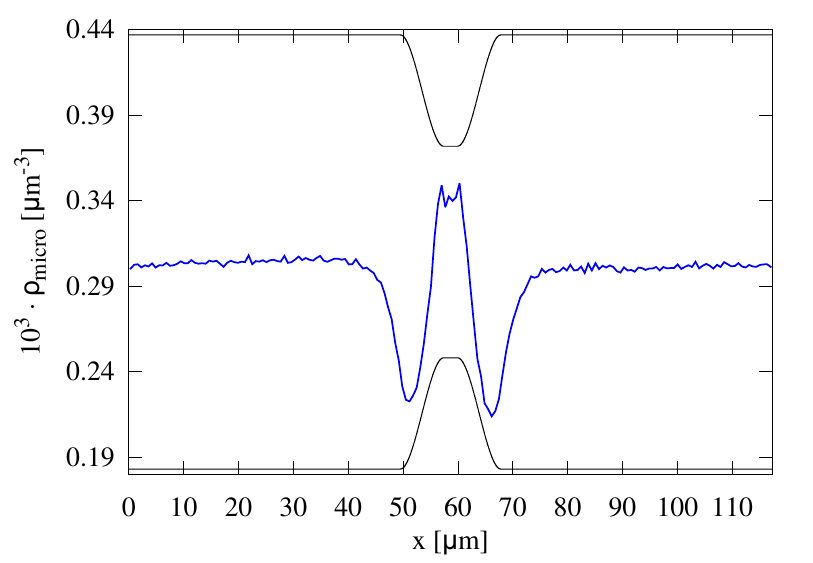}
	\end{minipage}
	c)	
	\begin{minipage}{0.3\textwidth}
		\centering
		\includegraphics[width=\textwidth]{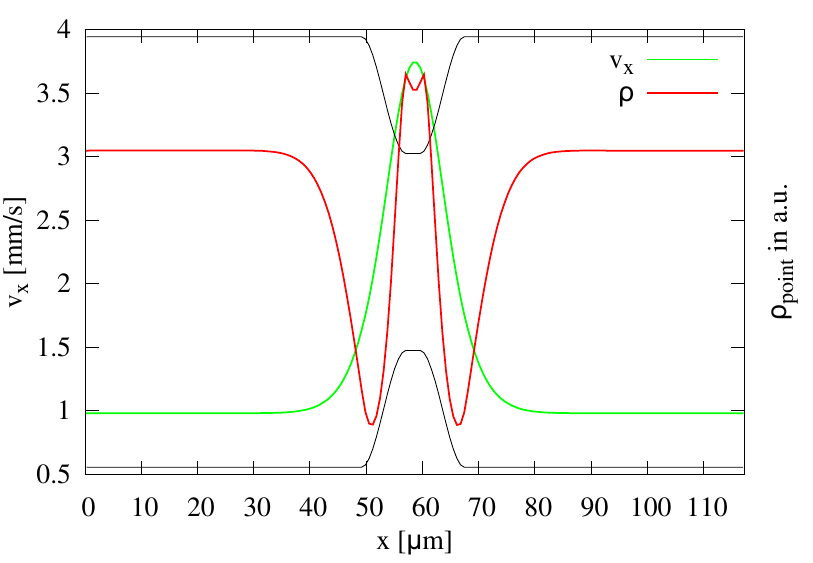}	 
	\end{minipage}	
	\caption{Concentration as a function of the axial position for a pure RBC suspension (a) and for a pure microparticle suspension (b). Despite very different particle properties and mean concentrations both profiles are remarkably similar.
	(c) The axial flow velocity on the centerline for the pure liquid without thermal fluctuations (green curve). The concentration distribution for a train of passive tracer particles following this velocity profile (red curve) well reproduces the concentration profiles of the single-component suspensions in (a) and (b). The channel geometry is illustrated by the black line.}
	\label{FIG:rhoX_RBC_Point}
\end{figure*}

In order to assess more quantitatively a possible influence of the constriction, we average the 2D concentration profiles of figure~\ref{FIG:map} over the radial position to obtain one-dimensional axial concentration profiles.
To discriminate between the generic influence of the constriction and the influence of margination we start by considering single-component suspensions where margination is not present.
In figure~\ref{FIG:rhoX_RBC_Point} (a) we show the concentration profile for a pure RBC suspension which has been created by replacing all microparticles by RBCs in the system of figure~\ref{FIG:system}~(b).
Far from the constriction the concentration exhibits a constant plateau as expected in a straight channel.
Upon entering the constriction, the concentration first decreases and then rises again until a roughly constant plateau inside the constriction is reached. 
At the exit, the concentration decreases a second time and soon after recovers its straight channel level.
Removing the red blood cells from the system in figure~\ref{FIG:system} and leaving only the 18 microparticles leads to the concentration profile in figure~\ref{FIG:rhoX_RBC_Point}~(b).
The microparticle concentration profile looks remarkably similar to that of the pure RBCs pointing to a common generic origin.

%Due to their deformability the RBCs migrate towards and hence are located at the channel center. 
%Thus, the constriction influences the RBCs mainly through the changed velocity field. 

In order to explain these observations, we revert to the most simple model system: passive tracer particles flowing along the central axis through the constriction.
For this, we first extract the velocity profile from a simulation without any particles and thermal fluctuations (Fig. \ref{FIG:rhoX_RBC_Point} (c) green curve).
We then compute the position as a function of time for a train of tracer particles which precisely follow this velocity. 
Due to periodic boundary conditions the particles repeatedly pass the constriction allowing us to compute the time-averaged concentration profile, which is shown as the red line in figure~\ref{FIG:rhoX_RBC_Point}~(c).
The profile of this simple system looks almost the same as the ones from the more complex RBC and microparticle suspensions.
This furnishes a straightforward explanation for the concentration profiles: upon entering the constriction the fluid velocity rises due to mass conservation leading to an acceleration of a flowing particle. 
The acceleration increases the distance to the following particle and thus causes a concentration decrease.
Inside the constriction, this effect however is (over-)compensated by the reduced channel radius leading to a plateau which is slightly higher than the straight channel value.
At the exit of the constriction the opposite dynamics is observed. 
It remains remarkable that the concentration profile of the RBC suspension in figure~\ref{FIG:rhoX_RBC_Point}~(a) and the microparticle suspension in figure~\ref{FIG:rhoX_RBC_Point}~(b) can be well reproduced by passive tracer points.
The only exception is that the RBC densities in front of and behind the constriction are slightly asymmetric while those of the particle train are perfectly symmetric as expected under Stokes flow conditions.
Similar asymmetries in RBC concentration have also been observed by \cite{Vahidkhah_2016_stenosis} and (in planar channels) by \cite{Yazdani_2016}.
The most likely source for the asymmetry lies in the RBC deformability. 

%To investigate the influence of the constriction on the velocity field and hence on the particle density we evaluate the velocity on the centerline for the same system, but without any particles and without thermal fluctuations (Fig. \ref{FIG:rhoX_RBC_Point} (b) \textcolor{green}{green} curve). 
%Due to the decreasing radius and mass conservation the velocity increases in front of and decreases behind the constriction. 
%In the picture of several particles the increase of the density can be explained by additional fluid that moves from higher radii into the space between the particles. The additional fluid increases the relative distance between the particles at the site in front of the constriction. Behind the constriction the effect applies vice versa. In the picture of a single point particle the additional fluid causes the increasing velocity in front of the constriction in Fig \ref{FIG:rhoX_RBC_Point} (b). Due to the increased velocity the point particle moves faster in front of the constriction and hence dwells less time in the regarding bins. In addition, the decreasing volume due to the decreasing radius inside the constriction overlaps the effect of the increased velocity and causes the density to increase. 

%Nevertheless, the axial density assuming a point particle agrees surprisingly well with the RBCs and hence provides a good explanation for the RBC density profile in Fig. \ref{FIG:rhoX_RBC_Point} (a). 

\subsection*{Clustering of microparticles at constrictions in mixed suspensions}

\begin{figure*}
	\centering
	%	a)
	%	\begin{minipage}{0.47\textwidth}
	%		\includegraphics[width=\textwidth]{images/xdistr_part1.pdf}
	%	\end{minipage}
	%	b)
	a)
	\begin{minipage}{0.46\textwidth}
		\includegraphics[width=\textwidth]{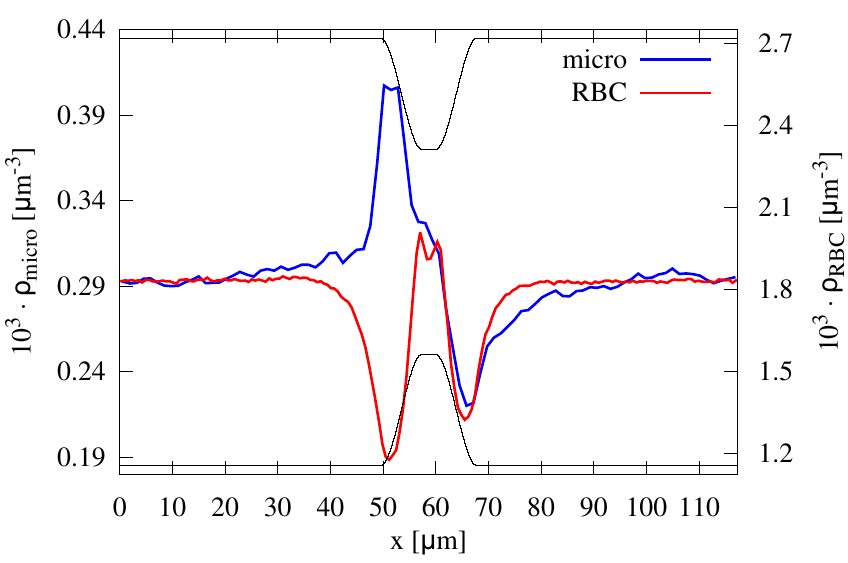}
	\end{minipage}
	b)
	\begin{minipage}{0.46\textwidth}
		\includegraphics[width=\textwidth]{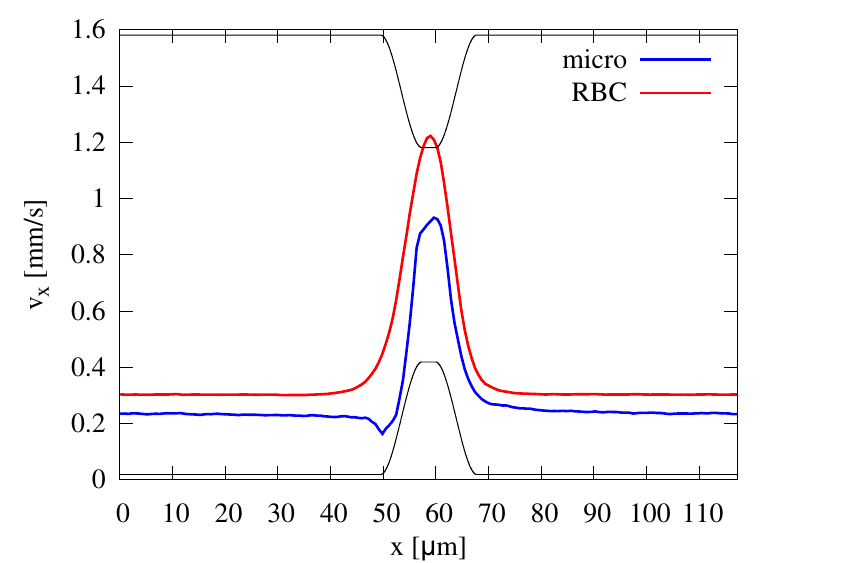}
	\end{minipage}
	c)
	\begin{minipage}{0.7\textwidth}
		\includegraphics[width=\textwidth]{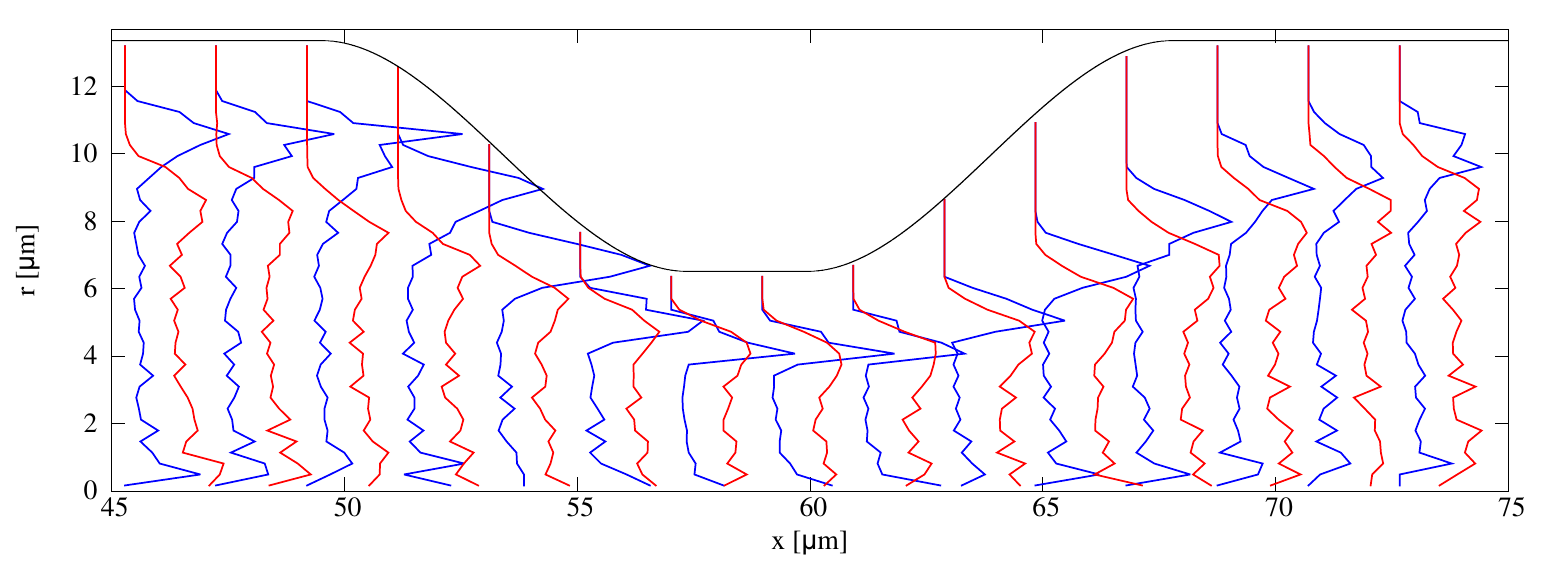}
	\end{minipage}
	\caption{(a) Concentration and (b) velocity as a function of axial position for the mixed suspension with Ht=16.5\% shown in figure~\ref{FIG:system}~(b). The RBCs (red curve) and the microparticles (blue curve) exhibit precisely opposite behavior in front of the constriction. While the RBCs decrease their concentration, the concentration of the microparticles significantly increases. c) Concentration as a function of the radial position for various sites along the constriction. Inside the constriction the microparticles are located at same radial positions as the RBCs. The channel geometry is illustrated by the black line.}
	\label{FIG:rhoX_stiff}
\end{figure*}

We now consider the mixed RBC/microparticle suspension with Ht=16.5\% as shown in figure~\ref{FIG:system}~(b).
Due to their position in the channel center the RBCs in this system are not affected by the small number of microparticles and exhibit the same behavior as in the pure RBC system (Fig. \ref{FIG:rhoX_RBC_Point}~(a)) as demonstrated by the red curve in figure~\ref{FIG:rhoX_stiff}~(a).
In stark contrast, the microparticle concentration shown by the blue line in figure~\ref{FIG:rhoX_stiff}~(a) is completely changed by the presence of the red blood cells. 
Instead of a decrease, the concentration in front of the constriction now exhibits a pronounced increase.
At the exit of the constriction the same dip in the concentration appears as seen before for the pure microparticle suspension in figure~\ref{FIG:rhoX_RBC_Point}~(c).

The increase of the microparticle concentration in front of the constriction, which is our main result, can only be explained by the interplay of margination and constriction. 
Due to their deformability the RBCs migrate to the channel center and are thus able to pass relatively unhindered through the constriction.
The dense central plug of RBCs however acts as a barrier for the microparticles.
Arriving on a marginated position at the channel walls, in order to pass the constriction, they need to penetrate the RBC plug.
The collisions with the RBCs counteract this motion resulting in a longer dwell time in front of the constriction.
This causes the peak in the microparticle concentration.

Indeed, this mechanism is demonstrated by the radial concentration profiles in figure \ref{FIG:rhoX_stiff}~(c).
Before and after the constriction the microparticles are clearly located inside the cell-free layer well separated from the RBCs.
Entering the constriction the particles move towards the center and the particle peak starts to overlap with the broad RBC profile.
Finally, inside the constricted part the microparticles and RBCs are located at similar radial positions, although the microparticles do retain a near-wall position in agreement with the planar situation investigated in \cite{Yazdani_2016} (although the authors did not observe the increase in concentration before the constriction).
Regarding the RBCs themselves, we find that the cell-free layer thickness transiently increases at the exit of the constriction as in \cite{Yazdani_2016, Vahidkhah_2016_stenosis}. 
The pausing in front of the constriction also reflects itself in the microparticle velocity in figure \ref{FIG:rhoX_stiff}~(b).
Whereas the RBC velocity monotonously increases ahead the constriction, the microparticle velocity exhibits a clear dip.
%Furthermore, the standard deviation illustrated by the dotted lines becomes larger near the dip in contrast to that of the RBCs.
%Both results reflect a broader distribution of microparticle velocity and the deceleration ahead the constriction.

Thus, the effect of the same constriction is directly opposite for RBCs and marginated microparticles: while the RBCs are diluted, the concentration of the microparticles considerably increases in front of the constriction.

\subsection*{Influence of hematocrit}
%
% Here also lower Ht
%

\begin{figure*}
	\centering
	a)
	\begin{minipage}{0.46\textwidth}
		\includegraphics[width=\textwidth]{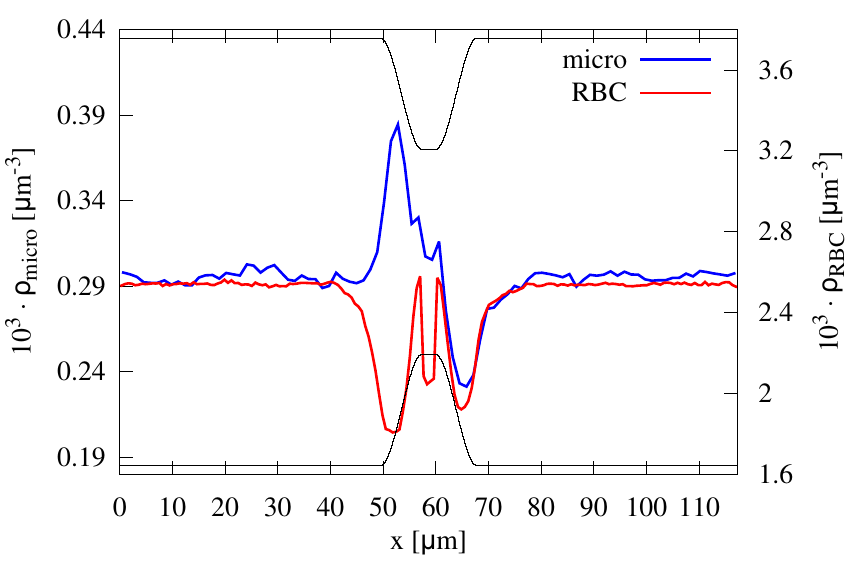}
	\end{minipage}
	b)
	\begin{minipage}{0.46\textwidth}
		\includegraphics[width=\textwidth]{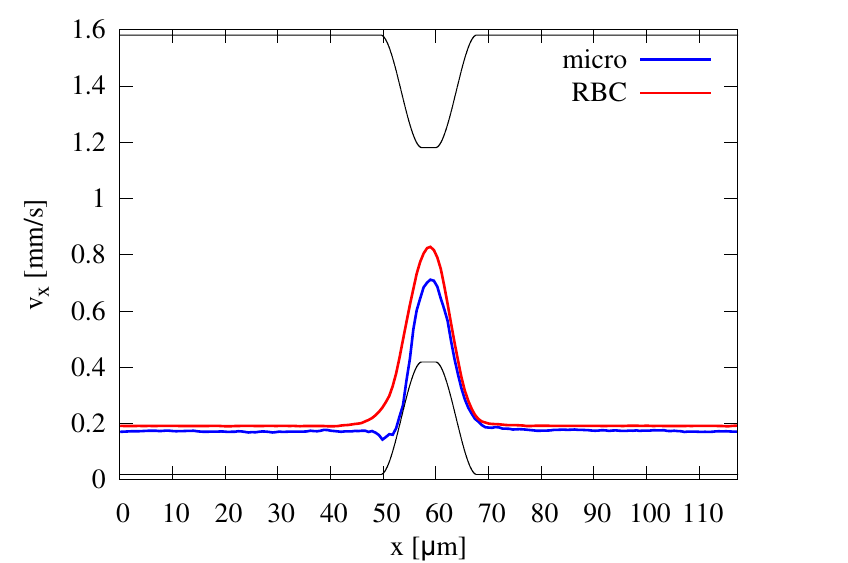}
	\end{minipage}
	c)
	\begin{minipage}{0.7\textwidth}
		\includegraphics[width=\textwidth]{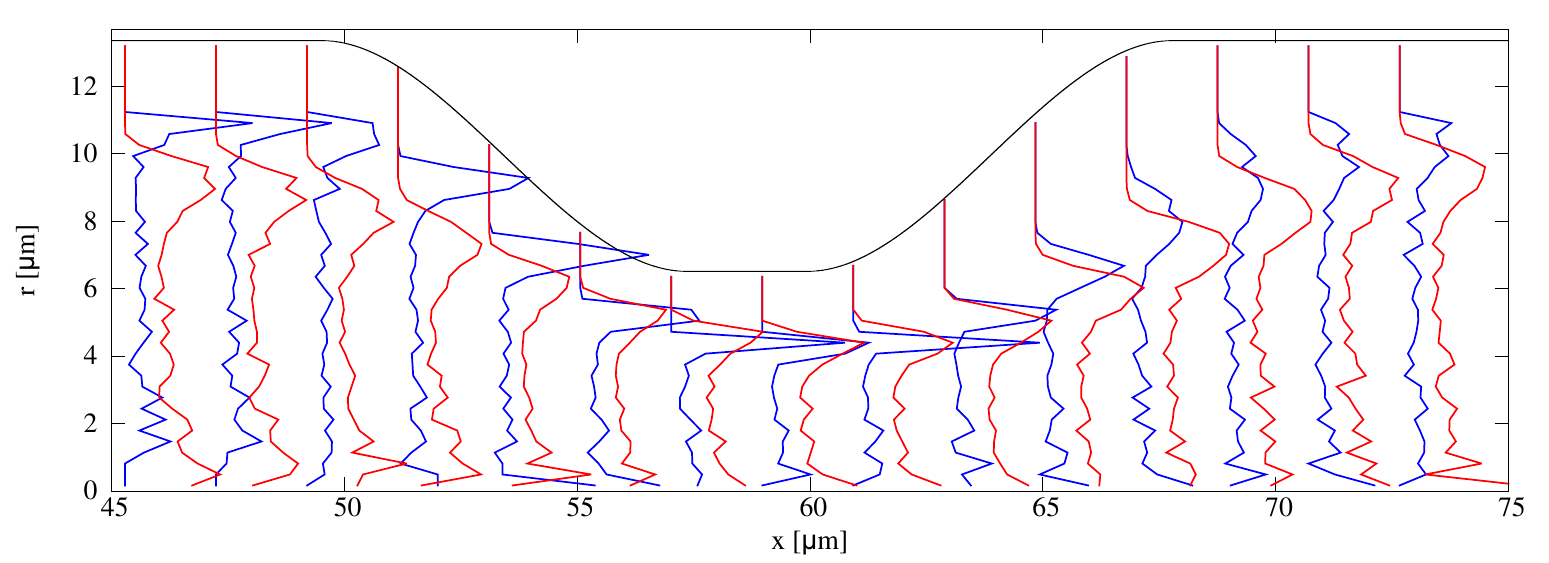}
	\end{minipage}
	\caption{(a) Concentration and (b) velocity as a function of axial position for a mixed suspension with hematocrit Ht=\num{23.7}\%. (c) Concentration as a function of the radial position for various sites along the constriction. The RBCs (red curve) exhibit the same behavior as for Ht=\num{16.5}\% in figure~\ref{FIG:rhoX_stiff} except  that the concentration inside the constriction is lower than in the main channel. The microparticles cluster in roughly the same amount ahead of the constriction. The channel geometry is illustrated by the black line.}
	\label{FIG:rhoX_Ht}
\end{figure*}

By increasing the number of RBCs in the system to $\num{151}$, we investigate the influence of a higher hematocrit Ht=23.7\% as shown in figure~\ref{FIG:system}~(c).
The axial concentration for Ht = \num{23.7}\% in figure \ref{FIG:rhoX_Ht}~(a) proofs that microparticle clustering also appears at higher Ht.
However, differences occur in RBC concentration: in contrast to the lower Ht in figure \ref{FIG:rhoX_stiff}~(a), RBCs are diluted inside the constriction compared to the main channel.
This effect is not captured by our simple isolated particle model in figure~\ref{FIG:rhoX_RBC_Point}~(c) showing the importance of collective RBC-RBC interactions at higher Ht.

In figure \ref{FIG:rhoX_Ht}~(c) the radial concentration profiles are shown for Ht = 23.7\%.
The dipping of microparticles into the RBC plug is present here in the same manner as for Ht = 16.5\% in figure \ref{FIG:rhoX_stiff}~(c).
While the velocity shown in figure \ref{FIG:rhoX_Ht}~(b) is in general lower due to the larger amount of cells, a dip in the microparticle velocity ahead the constriction is again observed.

The results for Ht=23.7\% suggest that an upper boundary for the extent of microparticle clustering exists.
A possible explanation may be provided by the discrete motion of the RBCs through the constriction\cite{Vahidkhah_2016_stenosis}.
Due to the limited space inside the constriction, even for higher Ht the number of simultaneously passing RBCs is not strongly increased.
As a consequence the probability for microparticles to find a gap and thus the clustering effect is nearly the same at both hematocrits.

\subsection*{Influence of constriction properties}

To investigate the influence of the constriction properties we conduct one simulation with a longer constriction, one with a narrower constriction and a third one with a longer (smoother) transition zone at Ht=16.5\%.

\begin{figure*}
	a)
	\begin{minipage}{0.3\textwidth}
		\centering
		\includegraphics[width=\textwidth]{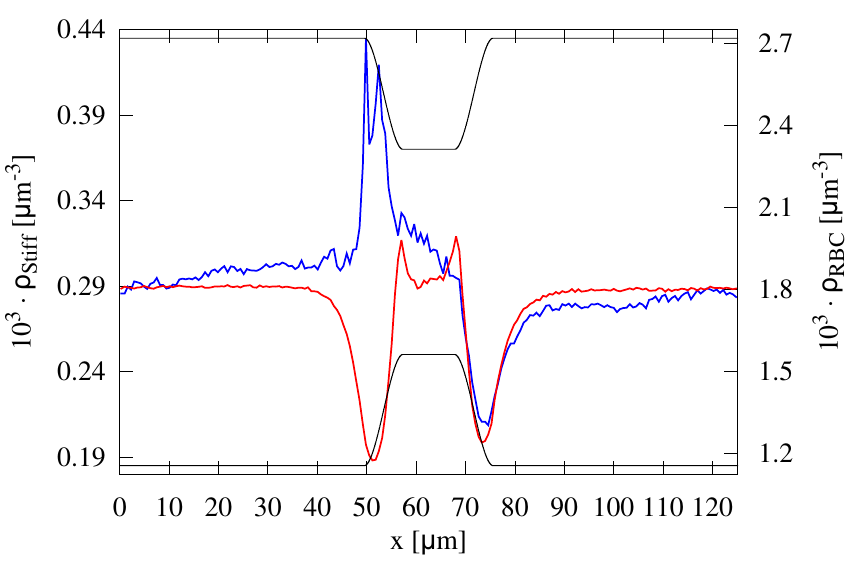}
	\end{minipage}
	b)
	\begin{minipage}{0.3\textwidth}
		\centering
		\includegraphics[width=\textwidth]{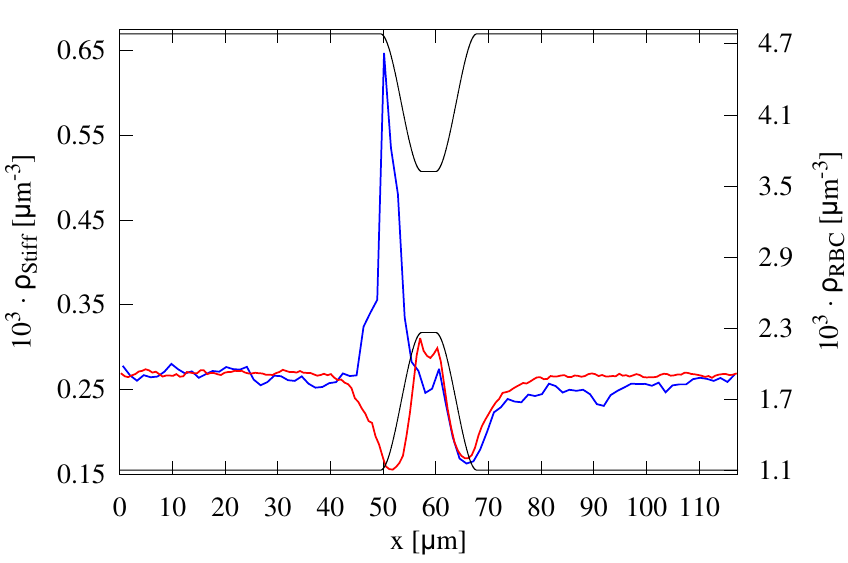}	
	\end{minipage}
	c)
	\begin{minipage}{0.3\textwidth}
		\centering
		\includegraphics[width=\textwidth]{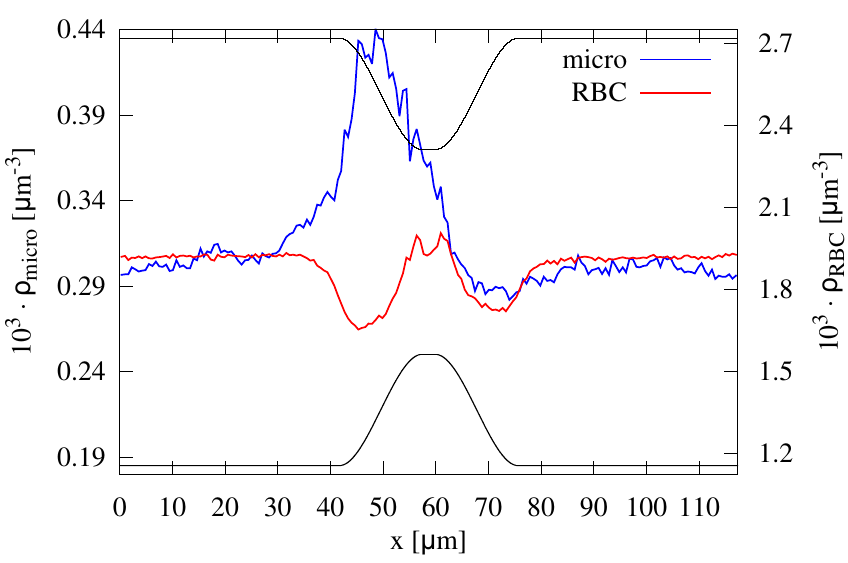}
	\end{minipage}		
	\caption{Concentration as function of the axial position for the RBCs (red curve) and microparticles (blue curve) for (a) a four times longer constriction, (b) a narrower constriction with a radii ratio equal 3/8, and (c) a twofold longer transition zone. The channel geometry is illustrated by the black line.}
	\label{FIG:rhoX_geometry}
\end{figure*}

In the first simulation, the constricted part is four times longer than in figure~\ref{FIG:system}~(b). 
The observed axial concentration distribution is shown in figure~\ref{FIG:rhoX_geometry} (a). 
The microparticle concentration increases in front of the constriction in the same manner as for the short constriction (Fig. \ref{FIG:rhoX_stiff}). Inside the constriction the microparticle concentration decreases slowly. 
We thus conclude that the clustering effect is independent of the length of the constriction.

The narrower constriction is chosen such that the ratio of the radii of the constricted part and the straight part equals 3/8 instead of 1/2 before. 
The increase in the microparticle concentration is significantly stronger (Fig. \ref{FIG:rhoX_geometry} (b)) compared to figure~\ref{FIG:rhoX_stiff}.
The explanation is straightforward: to pass the constriction a microparticle now has to migrate closer to the centerline. 
As explained before the collisions with the RBCs counteract this motion leading to a longer dwell time in front of the constriction. 

Finally, we double the length of the transition zone between the main cylinder and the constricted part leading to a smoother transition.
The observed concentration profile of the RBCs in figure~\ref{FIG:rhoX_geometry} (c) is similar to that in figure~\ref{FIG:rhoX_stiff}, except that the decrease ahead and behind the constriction is less pronounced.
This difference can be reproduced by the point particle method which shows the same changes in the concentration profile (data not shown).
%Due to the more weakly varying boundary the point at which the velocity starts to increase is closer to the beginning of the constriction. 
%Hence, the concentration decrease due to the increased velocity is earlier (over-)compensated by the decreasing volume than in the system in figure~\ref{FIG:system}.
Importantly, the amount of microparticle clustering is not influenced by the smoother transition. 
%Nevertheless, we note that the decrease in microparticle concentration vanishes.

These results lead us to suggest that the radius of the constriction is the important parameter that determines the amount of microparticle clustering: the narrower the constriction the stronger the clustering. However, microparticle clustering is independent of constriction length and transition zone. Below we show that the clustering in front of the constriction occurs for ellipsoidal microparticles, as well.

\subsection*{Influence of particle shape and size}
%\subsection*{Clustering of ellipsoidal microparticles}

%
% here also larger spherical particles
%

\begin{figure}[h]
	\centering
	\begin{minipage}{0.5\textwidth}
		\includegraphics[width=\textwidth]{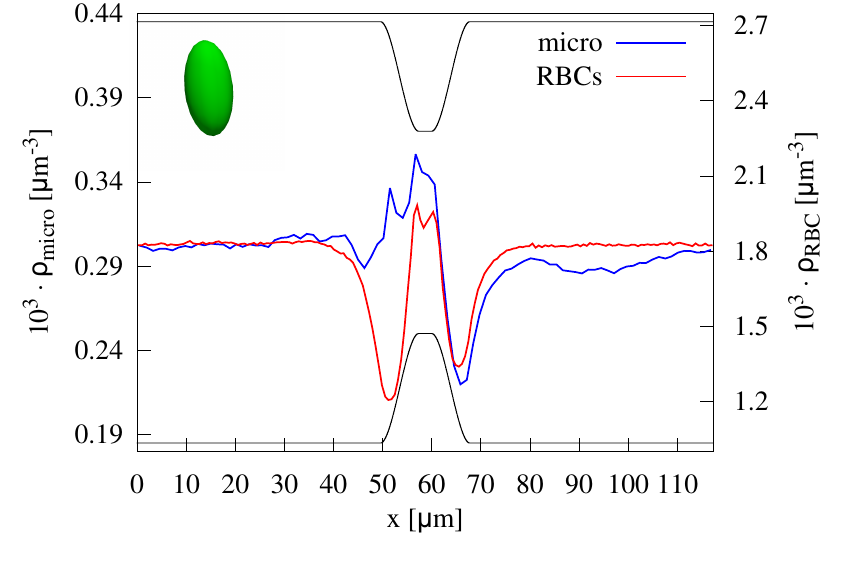}
	\end{minipage}
	\caption{Center-of-mass density depending on the axial position for the RBCs (red curve) and the stiff ellipsoids (blue curve) in a red blood cell suspension flowing through a constriction. The ellipsoidal microparticles cluster, but not in the same amount as the spherical microparticles. An image of an ellipsoid is inserted. The channel geometry is illustrated by the black line.}
	\label{FIG:rhoX_ellipsoids}
\end{figure}

Finally, we investigate the behavior of stiff ellipsoids to study a possible influence of the particle shape and size. 
For this, we use prolate ellipsoids with a ratio of two between the long and the short axis (Fig. \ref{FIG:rhoX_ellipsoids} inset). 
Along the short axis the ellipsoid diameter equals the diameter of the spheres used before. 
%%Since no clustering of ellipsoids occurred for the small channel with length $\SI{70}{\mu m}$ (data not shown) 
%The channel and constriction radii are equal to those used in figure~\ref{FIG:system}~(b), but we employ a longer channel with length $\SI{165}{\mu m}$. 
We find that also ellipsoids exhibit clustering in front of the constriction as shown in Fig. \ref{FIG:rhoX_ellipsoids}. 
Compared to the spherical microparticles the increase in concentration ahead of the constriction is however less pronounced. 
The slightly reduced clustering of the ellipsoids can be explained by the increased size of the ellipsoids along one principal axis. 
Due to this increased size the ellipsoids can dip more easily into the RBC plug and migrate towards the channel center. 
Nevertheless, the ellipsoid data suggests that clustering in front of constriction is a robust effect, which is qualitatively independent of the particle shape.

\section*{Conclusion}

% brief revisit and importance for the filed
We investigated the axial concentration of red blood cells and microparticles in a mixed suspension flowing through a constricted cylindrical microchannel as a model system for a stenosis in the microcirculation.
We found that the constriction causes a pronounced peak in the microparticle concentration right in front of the constriction. 
This clustering of microparticles is explained by the marginated near-wall position, on which the microparticles arrive at the constriction and the corresponding more central position of the RBCs.
As a consequence, microparticle passage through the constriction is severely hindered by the dense plug of RBCs occupying the channel center.
Hence, the microparticles get stuck in front of the constriction until they find a suitable gap among the RBCs leading to a prolonged dwell time, which in the end causes the increased concentration.
Interestingly, the red blood cells themselves behave directly opposite showing a dip in concentration before the constriction.
Furthermore, we investigated the dependency on hematocrit, stenosis, particle size, and shape.
The clustering effect is qualitatively robust.% and independent of the precise shape of the constriction and/or the microparticles.
%Despite the radius of the constriction the properties of the constriction and the particle shape have slight influences on the clustering \SG{??}

%Far from the constriction the marginated particles concentrate near the channel wall, whereas the RBCs concentrate near the channel center. Approaching the constriction the marginated particles have to migrate towards the channel center. Due to the RBCs in the center the marginated particles are pulled back by collisions with the RBCs.  We provided an explanation of the clustering by highlighting the necessity of margination by removing the red blood cells from the system. Without the red blood cells the clustering vanishes.
%Varying the radius and length of the constriction leads us to suggest that the radius of the constriction is the important factor for the amount of clustering. \TODO{Influence of Transition?}

Our discovery of microparticle clustering will be especially relevant for biochemically active particles such as drug delivery agents, where the zone right in front of constrictions may become a hot spot of biochemical activity.
We furthermore expect that similar clustering effects will also be present for other particles undergoing margination, such as blood platelets, which may increase the likelihood of thrombus formation ahead of vessel stenoses \cite{Fogelson_2015, Spann_2016}.

% impact on field + recommendations future work
%Our studies suggests that the clustering could be found in vivo and in vitro. It remains the task of future research to support the clustering in front of constriction by experiments.
%In our simulations we found the clustering for marginated stiff spherical particles. Besides for stiff spherical particles, the clustering should also occur for other particles undergoing margination, such as white blood cells or blood platelets. 
%Hence, it could be the basis for medical solutions to deal with a pathological narrowing. Inserting drugs into the blood circulatory margination and a pathological narrowing cause an increased density in front of the narrowing. Assuming a biochemical reaction rate directly proportional to the local density the clustering enhances the effect of the applied drugs at the site of the narrowing. Consequently, it is of great importance for medical applications. Also the penetration rate into the tissue of the white blood cells, which generally deal with pathological changes, may increase in front of the constriction.

\section*{Supplementary Materials}
\label{SI}

See supplementary materials for influence of the total channel length.
Furthermore, we provide the velocity and axial concentration profiles along the constriction for the simulations with different stenosis properties and with ellipsoidal microparticles.

\section*{Acknowledgements}

The authors thank the Gauss Centre for Supercomputing e.V. for providing
computing time on the GCS Supercomputer SuperMUC at the Leibniz
Supercomputing Centre.

\noindent This work was supported by the Volkswagen Foundation.

\newpage
\bibliography{./paper.bib}
\end{document}

%% file: materials_methods.tex
We perform our simulations by means of a 3D Lattice-Boltzmann method (LBM) implemented in the simulation package ESPResSo \cite{Limbach2006,Roehm2012,Arnold2013} extended by an immersed boundary method (IBM) algorithm \cite{Peskin2002,Mittal2005}. 
The LBM uses the spatially and temporally discretized Boltzmann equation to calculate fluid behavior \cite{Succi2001, Duenweg2008, Aidun2010}. 
The simulation package ESPResSo enables us to add thermal fluctuations  with a temperature $T=\SI{300}{K}$. 
The fluid in which the particles are suspended is chosen to mimic blood plasma with a density of $\SI{1000}{kg/m^3}$ and a viscosity of $\SI{1.2e-3}{Pa s}$.

Both RBCs and microparticles are realized via the IBM. 
The RBC diameter is chosen to be 7.82${\mu m}$.
The red blood cell membrane is modeled as an infinitely thin elastic sheet which is discretized by a set of nodes that are treated as Lagrangian points flowing with the fluid and that are connected by triangles. 
The forces between the fluid and the membrane nodes are calculated every time step by interpolating between the fluid nodes using an eight-point stencil. 
For the elastic properties we use the Skalak model \cite{BarthesBiesel2011,Freund2013,Farutin2014} with a shear modulus $k_S = \SI{5e-6}{N/m}$ and an area dilatation modulus $k_A = 100 k_S$. 
Bending forces are computed via the Helfrich model with a bending modulus $k_B = \SI{2e-19}{Nm}$ \cite{FreundRev2014, Guckenberger2016} using the bending algorithm denominated method A in \cite{Guckenberger2016}. 
We note that, while this rather simplistic algorithm is clearly not the most accurate method, it should be sufficient for the present purpose where the major focus is on the behavior of the microparticles, not the RBCs themselves.
In addition, we use an empirical volume conservation potential \cite{Krueger2016}. 

The spherical microparticles have a diameter of $a=3.2{\mu m}$ and are thus in a range for which strong margination was recently reported  \cite{Mueller2014}.
Numerically, they follow the same model as the RBCs, but with $\num{1000}$ times higher shear and bending moduli. In addition, they are stiffened by a 3D lattice inside the particle with lattice constant of one LBM grid cell. 
The grid is connected to the shell by elastic springs and a bending potential. 
The stability of our particle model is documented in \cite{Gekle2016}. 
A soft-sphere potential with cut-off radius of one grid cell between the particles and between the wall and a particle ensures numerical stability.

The constriction is modeled by two straight cylinders with different radii and a transition zone between them. 
To provide a smooth transition the radius in the transition zone follows a third order polynomial such that the function describing the boundary and its first derivative are continuous.  
To control the flow rate a pressure gradient is applied.
%that drives the flow is chosen such that the centerline velocity in the main cylinder is about $v = \SI{1}{mm/s}$. 
The Reynolds number calculated from the centerline velocity, the RBC radius $R_{RBC}$, and the kinematic viscosity of the fluid $\nu$ is $Re = \frac{R_{RBC} \cdot v}{\nu} = \mathcal{O}(10^{-3})$. The time step is chosen to $0.06{\mu s}$.% \SG{give time step}

At the beginning of the simulation the RBCs and the microparticles are inserted with a regular pattern. 
The system is then run until all remnants of this regular initial configuration have vanished and a steady state is achieved. 
Concentration profiles are obtained by time-averaging over a rather long part of the simulations ($\sim 90\%$ of the total simulation time) starting from these random configurations. The systems are simulated over about $10{s}$.

%To get an overview of the marginated state we evaluate the density depending on the axial, i.e. in flow direction, and the radial position. 
The axial and radial concentration is calculated by dividing the system in axial and radial bins. 
At every time step RBCs and microparticles are sorted into corresponding bins using their center-of-mass.
The concentration is calculated by dividing the averaged number of particles inside one bin by the volume of the bin. 

%To investigate the influence of a pathological narrowing of a blood vessel on the dynamics of synthesized particles suspended in blood we simulated in our main system (Fig. \ref{FIG:system}) 18 stiff spherical particles in a suspension of 108 RBCs in a Newtonian fluid with blood plasma viscosity $\SI{1.2e-3}{Pa s}$ (Fig. \ref{FIG:system}). Due to the low volume fraction of other constituents taking only the RBCs into account provides nearly full blood behavior.

%% file: paper_arxiv.bbl
\begin{thebibliography}{61}%
\makeatletter
\providecommand \@ifxundefined [1]{%
 \@ifx{#1\undefined}
}%
\providecommand \@ifnum [1]{%
 \ifnum #1\expandafter \@firstoftwo
 \else \expandafter \@secondoftwo
 \fi
}%
\providecommand \@ifx [1]{%
 \ifx #1\expandafter \@firstoftwo
 \else \expandafter \@secondoftwo
 \fi
}%
\providecommand \natexlab [1]{#1}%
\providecommand \enquote  [1]{``#1''}%
\providecommand \bibnamefont  [1]{#1}%
\providecommand \bibfnamefont [1]{#1}%
\providecommand \citenamefont [1]{#1}%
\providecommand \href@noop [0]{\@secondoftwo}%
\providecommand \href [0]{\begingroup \@sanitize@url \@href}%
\providecommand \@href[1]{\@@startlink{#1}\@@href}%
\providecommand \@@href[1]{\endgroup#1\@@endlink}%
\providecommand \@sanitize@url [0]{\catcode `\\12\catcode `\$12\catcode
  `\&12\catcode `\#12\catcode `\^12\catcode `\_12\catcode `\%12\relax}%
\providecommand \@@startlink[1]{}%
\providecommand \@@endlink[0]{}%
\providecommand \url  [0]{\begingroup\@sanitize@url \@url }%
\providecommand \@url [1]{\endgroup\@href {#1}{\urlprefix }}%
\providecommand \urlprefix  [0]{URL }%
\providecommand \Eprint [0]{\href }%
\providecommand \doibase [0]{http://dx.doi.org/}%
\providecommand \selectlanguage [0]{\@gobble}%
\providecommand \bibinfo  [0]{\@secondoftwo}%
\providecommand \bibfield  [0]{\@secondoftwo}%
\providecommand \translation [1]{[#1]}%
\providecommand \BibitemOpen [0]{}%
\providecommand \bibitemStop [0]{}%
\providecommand \bibitemNoStop [0]{.\EOS\space}%
\providecommand \EOS [0]{\spacefactor3000\relax}%
\providecommand \BibitemShut  [1]{\csname bibitem#1\endcsname}%
\let\auto@bib@innerbib\@empty
%</preamble>
\bibitem [{\citenamefont {Popel}\ and\ \citenamefont
  {Johnson}(2005)}]{Popel2005}%
  \BibitemOpen
  \bibfield  {author} {\bibinfo {author} {\bibfnamefont {A.~S.}\ \bibnamefont
  {Popel}}\ and\ \bibinfo {author} {\bibfnamefont {P.~C.}\ \bibnamefont
  {Johnson}},\ }\href@noop {} {\bibfield  {journal} {\bibinfo  {journal}
  {Annual review of fluid mechanics}\ }\textbf {\bibinfo {volume} {37}},\
  \bibinfo {pages} {43} (\bibinfo {year} {2005})}\BibitemShut {NoStop}%
\bibitem [{\citenamefont {Misbah}\ and\ \citenamefont
  {Wagner}(2013)}]{Misbah2013}%
  \BibitemOpen
  \bibfield  {author} {\bibinfo {author} {\bibfnamefont {C.}~\bibnamefont
  {Misbah}}\ and\ \bibinfo {author} {\bibfnamefont {C.}~\bibnamefont
  {Wagner}},\ }\href {\doibase http://dx.doi.org/10.1016/j.crhy.2013.05.004}
  {\bibfield  {journal} {\bibinfo  {journal} {Comptes Rendus Physique}\
  }\textbf {\bibinfo {volume} {14}},\ \bibinfo {pages} {447 } (\bibinfo {year}
  {2013})},\ \bibinfo {note} {living fluids / Fluides vivants}\BibitemShut
  {NoStop}%
\bibitem [{\citenamefont {Gompper}\ and\ \citenamefont
  {Fedosov}(2015)}]{Gompper_2015}%
  \BibitemOpen
  \bibfield  {author} {\bibinfo {author} {\bibfnamefont {G.}~\bibnamefont
  {Gompper}}\ and\ \bibinfo {author} {\bibfnamefont {D.~A.}\ \bibnamefont
  {Fedosov}},\ }\href@noop {} {\bibfield  {journal} {\bibinfo  {journal} {WIREs
  Syst Biol Med}\ }\textbf {\bibinfo {volume} {8}},\ \bibinfo {pages} {157}
  (\bibinfo {year} {2015})}\BibitemShut {NoStop}%
\bibitem [{\citenamefont {Freund}(2014)}]{FreundRev2014}%
  \BibitemOpen
  \bibfield  {author} {\bibinfo {author} {\bibfnamefont {J.~B.}\ \bibnamefont
  {Freund}},\ }\href {\doibase 10.1146/annurev-fluid-010313-141349} {\bibfield
  {journal} {\bibinfo  {journal} {Annual Review of Fluid Mechanics}\ }\textbf
  {\bibinfo {volume} {46}},\ \bibinfo {pages} {67} (\bibinfo {year} {2014})},\
  \Eprint
  {http://arxiv.org/abs/http://dx.doi.org/10.1146/annurev-fluid-010313-141349}
  {http://dx.doi.org/10.1146/annurev-fluid-010313-141349} \BibitemShut
  {NoStop}%
\bibitem [{\citenamefont {Schmid-Schönbein}\ \emph {et~al.}(1980)\citenamefont
  {Schmid-Schönbein}, \citenamefont {Usami}, \citenamefont {Skalak},\ and\
  \citenamefont {Chien}}]{SchmidSchoenbein1980}%
  \BibitemOpen
  \bibfield  {author} {\bibinfo {author} {\bibfnamefont {G.~W.}\ \bibnamefont
  {Schmid-Schönbein}}, \bibinfo {author} {\bibfnamefont {S.}~\bibnamefont
  {Usami}}, \bibinfo {author} {\bibfnamefont {R.}~\bibnamefont {Skalak}}, \
  and\ \bibinfo {author} {\bibfnamefont {S.}~\bibnamefont {Chien}},\ }\href
  {\doibase http://dx.doi.org/10.1016/0026-2862(80)90083-7} {\bibfield
  {journal} {\bibinfo  {journal} {Microvascular Research}\ }\textbf {\bibinfo
  {volume} {19}},\ \bibinfo {pages} {45 } (\bibinfo {year} {1980})}\BibitemShut
  {NoStop}%
\bibitem [{\citenamefont {Eckstein}\ \emph {et~al.}(1988)\citenamefont
  {Eckstein}, \citenamefont {Tilles},\ and\ \citenamefont
  {III}}]{Eckstein1988}%
  \BibitemOpen
  \bibfield  {author} {\bibinfo {author} {\bibfnamefont {E.~C.}\ \bibnamefont
  {Eckstein}}, \bibinfo {author} {\bibfnamefont {A.~W.}\ \bibnamefont
  {Tilles}}, \ and\ \bibinfo {author} {\bibfnamefont {F.~J.~M.}\ \bibnamefont
  {III}},\ }\href {\doibase http://dx.doi.org/10.1016/0026-2862(88)90036-2}
  {\bibfield  {journal} {\bibinfo  {journal} {Microvascular Research}\ }\textbf
  {\bibinfo {volume} {36}},\ \bibinfo {pages} {31 } (\bibinfo {year}
  {1988})}\BibitemShut {NoStop}%
\bibitem [{\citenamefont {Jain}\ and\ \citenamefont {Munn}(2009)}]{Jain2009}%
  \BibitemOpen
  \bibfield  {author} {\bibinfo {author} {\bibfnamefont {A.}~\bibnamefont
  {Jain}}\ and\ \bibinfo {author} {\bibfnamefont {L.~L.}\ \bibnamefont
  {Munn}},\ }\href@noop {} {\bibfield  {journal} {\bibinfo  {journal} {PLoS
  One}\ }\textbf {\bibinfo {volume} {4}},\ \bibinfo {pages} {e7104} (\bibinfo
  {year} {2009})}\BibitemShut {NoStop}%
\bibitem [{\citenamefont {Charoenphol}\ \emph {et~al.}(2010)\citenamefont
  {Charoenphol}, \citenamefont {Huang},\ and\ \citenamefont
  {Eniola-Adefeso}}]{Charoenphol2010}%
  \BibitemOpen
  \bibfield  {author} {\bibinfo {author} {\bibfnamefont {P.}~\bibnamefont
  {Charoenphol}}, \bibinfo {author} {\bibfnamefont {R.~B.}\ \bibnamefont
  {Huang}}, \ and\ \bibinfo {author} {\bibfnamefont {O.}~\bibnamefont
  {Eniola-Adefeso}},\ }\href {\doibase
  http://dx.doi.org/10.1016/j.biomaterials.2009.11.007} {\bibfield  {journal}
  {\bibinfo  {journal} {Biomaterials}\ }\textbf {\bibinfo {volume} {31}},\
  \bibinfo {pages} {1392 } (\bibinfo {year} {2010})}\BibitemShut {NoStop}%
\bibitem [{\citenamefont {Lee}\ \emph {et~al.}(2013{\natexlab{a}})\citenamefont
  {Lee}, \citenamefont {Choi}, \citenamefont {Kopacz}, \citenamefont {Yun},
  \citenamefont {Liu},\ and\ \citenamefont {Decuzzi}}]{Lee2013}%
  \BibitemOpen
  \bibfield  {author} {\bibinfo {author} {\bibfnamefont {T.-R.}\ \bibnamefont
  {Lee}}, \bibinfo {author} {\bibfnamefont {M.}~\bibnamefont {Choi}}, \bibinfo
  {author} {\bibfnamefont {A.~M.}\ \bibnamefont {Kopacz}}, \bibinfo {author}
  {\bibfnamefont {S.-H.}\ \bibnamefont {Yun}}, \bibinfo {author} {\bibfnamefont
  {W.~K.}\ \bibnamefont {Liu}}, \ and\ \bibinfo {author} {\bibfnamefont
  {P.}~\bibnamefont {Decuzzi}},\ }\href@noop {} {\bibfield  {journal} {\bibinfo
   {journal} {Scientific reports}\ }\textbf {\bibinfo {volume} {3}} (\bibinfo
  {year} {2013}{\natexlab{a}})}\BibitemShut {NoStop}%
\bibitem [{\citenamefont {Chen}\ \emph {et~al.}(2013)\citenamefont {Chen},
  \citenamefont {Angerer}, \citenamefont {Napoleone}, \citenamefont
  {Reininger}, \citenamefont {Schneider}, \citenamefont {Wixforth},
  \citenamefont {Schneider},\ and\ \citenamefont {Alexander-Katz}}]{Chen2013}%
  \BibitemOpen
  \bibfield  {author} {\bibinfo {author} {\bibfnamefont {H.}~\bibnamefont
  {Chen}}, \bibinfo {author} {\bibfnamefont {J.~I.}\ \bibnamefont {Angerer}},
  \bibinfo {author} {\bibfnamefont {M.}~\bibnamefont {Napoleone}}, \bibinfo
  {author} {\bibfnamefont {A.~J.}\ \bibnamefont {Reininger}}, \bibinfo {author}
  {\bibfnamefont {S.~W.}\ \bibnamefont {Schneider}}, \bibinfo {author}
  {\bibfnamefont {A.}~\bibnamefont {Wixforth}}, \bibinfo {author}
  {\bibfnamefont {M.~F.}\ \bibnamefont {Schneider}}, \ and\ \bibinfo {author}
  {\bibfnamefont {A.}~\bibnamefont {Alexander-Katz}},\ }\href {\doibase
  http://dx.doi.org/10.1063/1.4833975} {\bibfield  {journal} {\bibinfo
  {journal} {Biomicrofluidics}\ }\textbf {\bibinfo {volume} {7}},\ \bibinfo
  {eid} {064113} (\bibinfo {year} {2013}),\
  http://dx.doi.org/10.1063/1.4833975}\BibitemShut {NoStop}%
\bibitem [{\citenamefont {Namdee}\ \emph {et~al.}(2013)\citenamefont {Namdee},
  \citenamefont {Thompson}, \citenamefont {Charoenphol},\ and\ \citenamefont
  {Eniola-Adefeso}}]{Namdee_2013}%
  \BibitemOpen
  \bibfield  {author} {\bibinfo {author} {\bibfnamefont {K.}~\bibnamefont
  {Namdee}}, \bibinfo {author} {\bibfnamefont {A.~J.}\ \bibnamefont
  {Thompson}}, \bibinfo {author} {\bibfnamefont {P.}~\bibnamefont
  {Charoenphol}}, \ and\ \bibinfo {author} {\bibfnamefont {O.}~\bibnamefont
  {Eniola-Adefeso}},\ }\href@noop {} {\bibfield  {journal} {\bibinfo  {journal}
  {Langmuir}\ }\textbf {\bibinfo {volume} {29}},\ \bibinfo {pages} {2530}
  (\bibinfo {year} {2013})}\BibitemShut {NoStop}%
\bibitem [{\citenamefont {Wang}\ \emph
  {et~al.}(2013{\natexlab{a}})\citenamefont {Wang}, \citenamefont {Mao},
  \citenamefont {Byler}, \citenamefont {Patel}, \citenamefont {Henegar},
  \citenamefont {Alexeev},\ and\ \citenamefont {Sulchek}}]{Wang2013}%
  \BibitemOpen
  \bibfield  {author} {\bibinfo {author} {\bibfnamefont {G.}~\bibnamefont
  {Wang}}, \bibinfo {author} {\bibfnamefont {W.}~\bibnamefont {Mao}}, \bibinfo
  {author} {\bibfnamefont {R.}~\bibnamefont {Byler}}, \bibinfo {author}
  {\bibfnamefont {K.}~\bibnamefont {Patel}}, \bibinfo {author} {\bibfnamefont
  {C.}~\bibnamefont {Henegar}}, \bibinfo {author} {\bibfnamefont
  {A.}~\bibnamefont {Alexeev}}, \ and\ \bibinfo {author} {\bibfnamefont
  {T.}~\bibnamefont {Sulchek}},\ }\href@noop {} {\bibfield  {journal} {\bibinfo
   {journal} {PLoS One}\ }\textbf {\bibinfo {volume} {8}},\ \bibinfo {pages}
  {e75901} (\bibinfo {year} {2013}{\natexlab{a}})}\BibitemShut {NoStop}%
\bibitem [{\citenamefont {Lee}\ \emph {et~al.}(2013{\natexlab{b}})\citenamefont
  {Lee}, \citenamefont {Choi}, \citenamefont {Kopacz}, \citenamefont {Yun},
  \citenamefont {Liu},\ and\ \citenamefont {Decuzzi}}]{Lee_2013_margination}%
  \BibitemOpen
  \bibfield  {author} {\bibinfo {author} {\bibfnamefont {T.-R.}\ \bibnamefont
  {Lee}}, \bibinfo {author} {\bibfnamefont {M.}~\bibnamefont {Choi}}, \bibinfo
  {author} {\bibfnamefont {A.~M.}\ \bibnamefont {Kopacz}}, \bibinfo {author}
  {\bibfnamefont {S.-H.}\ \bibnamefont {Yun}}, \bibinfo {author} {\bibfnamefont
  {W.~K.}\ \bibnamefont {Liu}}, \ and\ \bibinfo {author} {\bibfnamefont
  {P.}~\bibnamefont {Decuzzi}},\ }\href@noop {} {\bibfield  {journal} {\bibinfo
   {journal} {Sci. Rep.}\ }\textbf {\bibinfo {volume} {3}},\ \bibinfo {pages}
  {1} (\bibinfo {year} {2013}{\natexlab{b}})}\BibitemShut {NoStop}%
\bibitem [{\citenamefont {Fitzgibbon}\ \emph {et~al.}(2015)\citenamefont
  {Fitzgibbon}, \citenamefont {Spann}, \citenamefont {Qi},\ and\ \citenamefont
  {Shaqfeh}}]{Fitzgibbon_2015}%
  \BibitemOpen
  \bibfield  {author} {\bibinfo {author} {\bibfnamefont {S.}~\bibnamefont
  {Fitzgibbon}}, \bibinfo {author} {\bibfnamefont {A.~P.}\ \bibnamefont
  {Spann}}, \bibinfo {author} {\bibfnamefont {Q.~M.}\ \bibnamefont {Qi}}, \
  and\ \bibinfo {author} {\bibfnamefont {E.~S.~G.}\ \bibnamefont {Shaqfeh}},\
  }\href@noop {} {\bibfield  {journal} {\bibinfo  {journal} {Biophys J}\
  }\textbf {\bibinfo {volume} {108}},\ \bibinfo {pages} {2601} (\bibinfo {year}
  {2015})}\BibitemShut {NoStop}%
\bibitem [{\citenamefont {Migliorini}\ \emph {et~al.}(2002)\citenamefont
  {Migliorini}, \citenamefont {Qian}, \citenamefont {Chen}, \citenamefont
  {Brown}, \citenamefont {Jain},\ and\ \citenamefont {Munn}}]{Migliorini2002}%
  \BibitemOpen
  \bibfield  {author} {\bibinfo {author} {\bibfnamefont {C.}~\bibnamefont
  {Migliorini}}, \bibinfo {author} {\bibfnamefont {Y.}~\bibnamefont {Qian}},
  \bibinfo {author} {\bibfnamefont {H.}~\bibnamefont {Chen}}, \bibinfo {author}
  {\bibfnamefont {E.~B.}\ \bibnamefont {Brown}}, \bibinfo {author}
  {\bibfnamefont {R.~K.}\ \bibnamefont {Jain}}, \ and\ \bibinfo {author}
  {\bibfnamefont {L.~L.}\ \bibnamefont {Munn}},\ }\href {\doibase
  http://dx.doi.org/10.1016/S0006-3495(02)73948-9} {\bibfield  {journal}
  {\bibinfo  {journal} {Biophysical Journal}\ }\textbf {\bibinfo {volume}
  {83}},\ \bibinfo {pages} {1834 } (\bibinfo {year} {2002})}\BibitemShut
  {NoStop}%
\bibitem [{\citenamefont {Freund}(2007)}]{Freund2007}%
  \BibitemOpen
  \bibfield  {author} {\bibinfo {author} {\bibfnamefont {J.~B.}\ \bibnamefont
  {Freund}},\ }\href {\doibase http://dx.doi.org/10.1063/1.2472479} {\bibfield
  {journal} {\bibinfo  {journal} {Physics of Fluids}\ }\textbf {\bibinfo
  {volume} {19}},\ \bibinfo {eid} {023301} (\bibinfo {year} {2007}),\
  http://dx.doi.org/10.1063/1.2472479}\BibitemShut {NoStop}%
\bibitem [{\citenamefont {Crowl}\ and\ \citenamefont
  {Fogelson}(2011)}]{Crowl2011}%
  \BibitemOpen
  \bibfield  {author} {\bibinfo {author} {\bibfnamefont {L.}~\bibnamefont
  {Crowl}}\ and\ \bibinfo {author} {\bibfnamefont {A.~L.}\ \bibnamefont
  {Fogelson}},\ }\href@noop {} {\bibfield  {journal} {\bibinfo  {journal}
  {Journal of fluid mechanics}\ }\textbf {\bibinfo {volume} {676}},\ \bibinfo
  {pages} {348} (\bibinfo {year} {2011})}\BibitemShut {NoStop}%
\bibitem [{\citenamefont {Kumar}\ and\ \citenamefont
  {Graham}(2011)}]{Kumar2011}%
  \BibitemOpen
  \bibfield  {author} {\bibinfo {author} {\bibfnamefont {A.}~\bibnamefont
  {Kumar}}\ and\ \bibinfo {author} {\bibfnamefont {M.~D.}\ \bibnamefont
  {Graham}},\ }\href {\doibase 10.1103/PhysRevE.84.066316} {\bibfield
  {journal} {\bibinfo  {journal} {Phys. Rev. E}\ }\textbf {\bibinfo {volume}
  {84}},\ \bibinfo {pages} {066316} (\bibinfo {year} {2011})}\BibitemShut
  {NoStop}%
\bibitem [{\citenamefont {Tokarev}\ \emph {et~al.}(2011)\citenamefont
  {Tokarev}, \citenamefont {Butylin}, \citenamefont {Ermakova}, \citenamefont
  {Shnol}, \citenamefont {Panasenko},\ and\ \citenamefont
  {Ataullakhanov}}]{Tokarev2011}%
  \BibitemOpen
  \bibfield  {author} {\bibinfo {author} {\bibfnamefont {A.}~\bibnamefont
  {Tokarev}}, \bibinfo {author} {\bibfnamefont {A.}~\bibnamefont {Butylin}},
  \bibinfo {author} {\bibfnamefont {E.}~\bibnamefont {Ermakova}}, \bibinfo
  {author} {\bibfnamefont {E.}~\bibnamefont {Shnol}}, \bibinfo {author}
  {\bibfnamefont {G.}~\bibnamefont {Panasenko}}, \ and\ \bibinfo {author}
  {\bibfnamefont {F.}~\bibnamefont {Ataullakhanov}},\ }\href {\doibase
  http://dx.doi.org/10.1016/j.bpj.2011.08.031} {\bibfield  {journal} {\bibinfo
  {journal} {Biophysical Journal}\ }\textbf {\bibinfo {volume} {101}},\
  \bibinfo {pages} {1835 } (\bibinfo {year} {2011})}\BibitemShut {NoStop}%
\bibitem [{\citenamefont {Zhao}\ and\ \citenamefont
  {Shaqfeh}(2011)}]{Zhao2011}%
  \BibitemOpen
  \bibfield  {author} {\bibinfo {author} {\bibfnamefont {H.}~\bibnamefont
  {Zhao}}\ and\ \bibinfo {author} {\bibfnamefont {E.~S.~G.}\ \bibnamefont
  {Shaqfeh}},\ }\href {\doibase 10.1103/PhysRevE.83.061924} {\bibfield
  {journal} {\bibinfo  {journal} {Phys. Rev. E}\ }\textbf {\bibinfo {volume}
  {83}},\ \bibinfo {pages} {061924} (\bibinfo {year} {2011})}\BibitemShut
  {NoStop}%
\bibitem [{\citenamefont {Zhao}\ \emph {et~al.}(2012)\citenamefont {Zhao},
  \citenamefont {Shaqfeh},\ and\ \citenamefont {Narsimhan}}]{Zhao2012}%
  \BibitemOpen
  \bibfield  {author} {\bibinfo {author} {\bibfnamefont {H.}~\bibnamefont
  {Zhao}}, \bibinfo {author} {\bibfnamefont {E.~S.~G.}\ \bibnamefont
  {Shaqfeh}}, \ and\ \bibinfo {author} {\bibfnamefont {V.}~\bibnamefont
  {Narsimhan}},\ }\href {\doibase http://dx.doi.org/10.1063/1.3677935}
  {\bibfield  {journal} {\bibinfo  {journal} {Physics of Fluids}\ }\textbf
  {\bibinfo {volume} {24}},\ \bibinfo {eid} {011902} (\bibinfo {year} {2012}),\
  http://dx.doi.org/10.1063/1.3677935}\BibitemShut {NoStop}%
\bibitem [{\citenamefont {Kumar}\ and\ \citenamefont
  {Graham}(2012)}]{Kumar2012}%
  \BibitemOpen
  \bibfield  {author} {\bibinfo {author} {\bibfnamefont {A.}~\bibnamefont
  {Kumar}}\ and\ \bibinfo {author} {\bibfnamefont {M.~D.}\ \bibnamefont
  {Graham}},\ }\href {\doibase 10.1103/PhysRevLett.109.108102} {\bibfield
  {journal} {\bibinfo  {journal} {Phys. Rev. Lett.}\ }\textbf {\bibinfo
  {volume} {109}},\ \bibinfo {pages} {108102} (\bibinfo {year}
  {2012})}\BibitemShut {NoStop}%
\bibitem [{\citenamefont {Fedosov}\ \emph {et~al.}(2012)\citenamefont
  {Fedosov}, \citenamefont {Fornleitner},\ and\ \citenamefont
  {Gompper}}]{Fedosov2012}%
  \BibitemOpen
  \bibfield  {author} {\bibinfo {author} {\bibfnamefont {D.~A.}\ \bibnamefont
  {Fedosov}}, \bibinfo {author} {\bibfnamefont {J.}~\bibnamefont
  {Fornleitner}}, \ and\ \bibinfo {author} {\bibfnamefont {G.}~\bibnamefont
  {Gompper}},\ }\href {\doibase 10.1103/PhysRevLett.108.028104} {\bibfield
  {journal} {\bibinfo  {journal} {Phys. Rev. Lett.}\ }\textbf {\bibinfo
  {volume} {108}},\ \bibinfo {pages} {028104} (\bibinfo {year}
  {2012})}\BibitemShut {NoStop}%
\bibitem [{\citenamefont {Freund}\ and\ \citenamefont
  {Shapiro}(2012)}]{Freund_2012}%
  \BibitemOpen
  \bibfield  {author} {\bibinfo {author} {\bibfnamefont {J.~B.}\ \bibnamefont
  {Freund}}\ and\ \bibinfo {author} {\bibfnamefont {B.}~\bibnamefont
  {Shapiro}},\ }\href@noop {} {\bibfield  {journal} {\bibinfo  {journal} {Phys.
  Fluids}\ }\textbf {\bibinfo {volume} {24}},\ \bibinfo {pages} {051904}
  (\bibinfo {year} {2012})}\BibitemShut {NoStop}%
\bibitem [{\citenamefont {Reasor}\ \emph {et~al.}(2013)\citenamefont {Reasor},
  \citenamefont {Mehrabadi}, \citenamefont {Ku},\ and\ \citenamefont
  {Aidun}}]{Reasor2013}%
  \BibitemOpen
  \bibfield  {author} {\bibinfo {author} {\bibfnamefont {D.~A.}\ \bibnamefont
  {Reasor}}, \bibinfo {author} {\bibfnamefont {M.}~\bibnamefont {Mehrabadi}},
  \bibinfo {author} {\bibfnamefont {D.~N.}\ \bibnamefont {Ku}}, \ and\ \bibinfo
  {author} {\bibfnamefont {C.~K.}\ \bibnamefont {Aidun}},\ }\href {\doibase
  10.1007/s10439-012-0648-7} {\bibfield  {journal} {\bibinfo  {journal} {Annals
  of Biomedical Engineering}\ }\textbf {\bibinfo {volume} {41}},\ \bibinfo
  {pages} {238} (\bibinfo {year} {2013})}\BibitemShut {NoStop}%
\bibitem [{\citenamefont {Kumar}\ \emph {et~al.}(2014)\citenamefont {Kumar},
  \citenamefont {Rivera},\ and\ \citenamefont {Graham}}]{Kumar2014}%
  \BibitemOpen
  \bibfield  {author} {\bibinfo {author} {\bibfnamefont {A.}~\bibnamefont
  {Kumar}}, \bibinfo {author} {\bibfnamefont {R.~G.~H.}\ \bibnamefont
  {Rivera}}, \ and\ \bibinfo {author} {\bibfnamefont {M.~D.}\ \bibnamefont
  {Graham}},\ }\href@noop {} {\bibfield  {journal} {\bibinfo  {journal}
  {Journal of Fluid Mechanics}\ }\textbf {\bibinfo {volume} {738}},\ \bibinfo
  {pages} {423} (\bibinfo {year} {2014})}\BibitemShut {NoStop}%
\bibitem [{\citenamefont {Fedosov}\ and\ \citenamefont
  {Gompper}(2014)}]{Fedosov_Gompper_2014}%
  \BibitemOpen
  \bibfield  {author} {\bibinfo {author} {\bibfnamefont {D.~A.}\ \bibnamefont
  {Fedosov}}\ and\ \bibinfo {author} {\bibfnamefont {G.}~\bibnamefont
  {Gompper}},\ }\href {\doibase 10.1039/C3SM52860J} {\bibfield  {journal}
  {\bibinfo  {journal} {Soft Matter}\ }\textbf {\bibinfo {volume} {10}},\
  \bibinfo {pages} {2961} (\bibinfo {year} {2014})}\BibitemShut {NoStop}%
\bibitem [{\citenamefont {M{\"u}ller}\ \emph {et~al.}(2014)\citenamefont
  {M{\"u}ller}, \citenamefont {Fedosov},\ and\ \citenamefont
  {Gompper}}]{Mueller2014}%
  \BibitemOpen
  \bibfield  {author} {\bibinfo {author} {\bibfnamefont {K.}~\bibnamefont
  {M{\"u}ller}}, \bibinfo {author} {\bibfnamefont {D.~A.}\ \bibnamefont
  {Fedosov}}, \ and\ \bibinfo {author} {\bibfnamefont {G.}~\bibnamefont
  {Gompper}},\ }\href@noop {} {\bibfield  {journal} {\bibinfo  {journal}
  {Scientific reports}\ }\textbf {\bibinfo {volume} {4}} (\bibinfo {year}
  {2014})}\BibitemShut {NoStop}%
\bibitem [{\citenamefont {Vahidkhah}\ \emph {et~al.}(2014)\citenamefont
  {Vahidkhah}, \citenamefont {Diamond},\ and\ \citenamefont
  {Bagchi}}]{Vahidkhah_2014}%
  \BibitemOpen
  \bibfield  {author} {\bibinfo {author} {\bibfnamefont {K.}~\bibnamefont
  {Vahidkhah}}, \bibinfo {author} {\bibfnamefont {S.~L.}\ \bibnamefont
  {Diamond}}, \ and\ \bibinfo {author} {\bibfnamefont {P.}~\bibnamefont
  {Bagchi}},\ }\href@noop {} {\bibfield  {journal} {\bibinfo  {journal}
  {Biophys J}\ }\textbf {\bibinfo {volume} {106}},\ \bibinfo {pages} {2529}
  (\bibinfo {year} {2014})}\BibitemShut {NoStop}%
\bibitem [{\citenamefont {Vahidkhah}\ and\ \citenamefont
  {Bagchi}(2015)}]{Vahidkhah_2015}%
  \BibitemOpen
  \bibfield  {author} {\bibinfo {author} {\bibfnamefont {K.}~\bibnamefont
  {Vahidkhah}}\ and\ \bibinfo {author} {\bibfnamefont {P.}~\bibnamefont
  {Bagchi}},\ }\href@noop {} {\bibfield  {journal} {\bibinfo  {journal} {Soft
  Matter}\ }\textbf {\bibinfo {volume} {11}},\ \bibinfo {pages} {2097}
  (\bibinfo {year} {2015})}\BibitemShut {NoStop}%
\bibitem [{\citenamefont {Henr\'{\i}quez~Rivera}\ \emph
  {et~al.}(2015)\citenamefont {Henr\'{\i}quez~Rivera}, \citenamefont {Sinha},\
  and\ \citenamefont {Graham}}]{Rivera2015}%
  \BibitemOpen
  \bibfield  {author} {\bibinfo {author} {\bibfnamefont {R.~G.}\ \bibnamefont
  {Henr\'{\i}quez~Rivera}}, \bibinfo {author} {\bibfnamefont {K.}~\bibnamefont
  {Sinha}}, \ and\ \bibinfo {author} {\bibfnamefont {M.~D.}\ \bibnamefont
  {Graham}},\ }\href {\doibase 10.1103/PhysRevLett.114.188101} {\bibfield
  {journal} {\bibinfo  {journal} {Phys. Rev. Lett.}\ }\textbf {\bibinfo
  {volume} {114}},\ \bibinfo {pages} {188101} (\bibinfo {year}
  {2015})}\BibitemShut {NoStop}%
\bibitem [{\citenamefont {M{\"u}ller}\ \emph {et~al.}(2016)\citenamefont
  {M{\"u}ller}, \citenamefont {Fedosov},\ and\ \citenamefont
  {Gompper}}]{Mueller2016}%
  \BibitemOpen
  \bibfield  {author} {\bibinfo {author} {\bibfnamefont {K.}~\bibnamefont
  {M{\"u}ller}}, \bibinfo {author} {\bibfnamefont {D.~A.}\ \bibnamefont
  {Fedosov}}, \ and\ \bibinfo {author} {\bibfnamefont {G.}~\bibnamefont
  {Gompper}},\ }\href {\doibase
  http://dx.doi.org/10.1016/j.medengphy.2015.08.009} {\bibfield  {journal}
  {\bibinfo  {journal} {Medical Engineering \& Physics}\ }\textbf {\bibinfo
  {volume} {38}},\ \bibinfo {pages} {2 } (\bibinfo {year} {2016})},\ \bibinfo
  {note} {micro and Nano Flows 2014 (MNF2014) - Biomedical Stream}\BibitemShut
  {NoStop}%
\bibitem [{\citenamefont {Gekle}(2016)}]{Gekle2016}%
  \BibitemOpen
  \bibfield  {author} {\bibinfo {author} {\bibfnamefont {S.}~\bibnamefont
  {Gekle}},\ }\href {\doibase http://dx.doi.org/10.1016/j.bpj.2015.12.005}
  {\bibfield  {journal} {\bibinfo  {journal} {Biophysical Journal}\ }\textbf
  {\bibinfo {volume} {110}},\ \bibinfo {pages} {514 } (\bibinfo {year}
  {2016})}\BibitemShut {NoStop}%
\bibitem [{\citenamefont {Kr{\"u}ger}(2016)}]{Krueger2016}%
  \BibitemOpen
  \bibfield  {author} {\bibinfo {author} {\bibfnamefont {T.}~\bibnamefont
  {Kr{\"u}ger}},\ }\href {\doibase 10.1007/s00397-015-0891-6} {\bibfield
  {journal} {\bibinfo  {journal} {Rheologica Acta}\ }\textbf {\bibinfo {volume}
  {55}},\ \bibinfo {pages} {511} (\bibinfo {year} {2016})}\BibitemShut
  {NoStop}%
\bibitem [{\citenamefont {Mehrabadi}\ \emph {et~al.}(2016)\citenamefont
  {Mehrabadi}, \citenamefont {Ku},\ and\ \citenamefont
  {Aidun}}]{Mehrabadi_2016}%
  \BibitemOpen
  \bibfield  {author} {\bibinfo {author} {\bibfnamefont {M.}~\bibnamefont
  {Mehrabadi}}, \bibinfo {author} {\bibfnamefont {D.~N.}\ \bibnamefont {Ku}}, \
  and\ \bibinfo {author} {\bibfnamefont {C.~K.}\ \bibnamefont {Aidun}},\
  }\href@noop {} {\bibfield  {journal} {\bibinfo  {journal} {Phys. Rev. E}\
  }\textbf {\bibinfo {volume} {93}},\ \bibinfo {pages} {023109} (\bibinfo
  {year} {2016})}\BibitemShut {NoStop}%
\bibitem [{\citenamefont {Spann}\ \emph {et~al.}(2016)\citenamefont {Spann},
  \citenamefont {Campbell}, \citenamefont {Fitzgibbon}, \citenamefont
  {Rodriguez}, \citenamefont {Cap}, \citenamefont {Blackbourne},\ and\
  \citenamefont {Shaqfeh}}]{Spann_2016}%
  \BibitemOpen
  \bibfield  {author} {\bibinfo {author} {\bibfnamefont {A.~P.}\ \bibnamefont
  {Spann}}, \bibinfo {author} {\bibfnamefont {J.~E.}\ \bibnamefont {Campbell}},
  \bibinfo {author} {\bibfnamefont {S.~R.}\ \bibnamefont {Fitzgibbon}},
  \bibinfo {author} {\bibfnamefont {A.}~\bibnamefont {Rodriguez}}, \bibinfo
  {author} {\bibfnamefont {A.~P.}\ \bibnamefont {Cap}}, \bibinfo {author}
  {\bibfnamefont {L.~H.}\ \bibnamefont {Blackbourne}}, \ and\ \bibinfo {author}
  {\bibfnamefont {E.~S.~G.}\ \bibnamefont {Shaqfeh}},\ }\href@noop {}
  {\bibfield  {journal} {\bibinfo  {journal} {Biophys J}\ }\textbf {\bibinfo
  {volume} {111}},\ \bibinfo {pages} {577} (\bibinfo {year}
  {2016})}\BibitemShut {NoStop}%
\bibitem [{\citenamefont {Vahidkhah}\ \emph {et~al.}(2016)\citenamefont
  {Vahidkhah}, \citenamefont {Balogh},\ and\ \citenamefont
  {Bagchi}}]{Vahidkhah_2016_stenosis}%
  \BibitemOpen
  \bibfield  {author} {\bibinfo {author} {\bibfnamefont {K.}~\bibnamefont
  {Vahidkhah}}, \bibinfo {author} {\bibfnamefont {P.}~\bibnamefont {Balogh}}, \
  and\ \bibinfo {author} {\bibfnamefont {P.}~\bibnamefont {Bagchi}},\
  }\href@noop {} {\bibfield  {journal} {\bibinfo  {journal} {Scientific
  Reports}\ }\textbf {\bibinfo {volume} {6}} (\bibinfo {year}
  {2016})}\BibitemShut {NoStop}%
\bibitem [{\citenamefont {Mountrakis}\ \emph {et~al.}(2013)\citenamefont
  {Mountrakis}, \citenamefont {Lorenz},\ and\ \citenamefont
  {Hoekstra}}]{Mountrakis2013}%
  \BibitemOpen
  \bibfield  {author} {\bibinfo {author} {\bibfnamefont {L.}~\bibnamefont
  {Mountrakis}}, \bibinfo {author} {\bibfnamefont {E.}~\bibnamefont {Lorenz}},
  \ and\ \bibinfo {author} {\bibfnamefont {A.~G.}\ \bibnamefont {Hoekstra}},\
  }\href {\doibase 10.1098/rsfs.2012.0089} {\bibfield  {journal} {\bibinfo
  {journal} {Interface Focus}\ }\textbf {\bibinfo {volume} {3}} (\bibinfo
  {year} {2013}),\ 10.1098/rsfs.2012.0089},\ \Eprint
  {http://arxiv.org/abs/http://rsfs.royalsocietypublishing.org/content/3/2/20120089.full.pdf}
  {http://rsfs.royalsocietypublishing.org/content/3/2/20120089.full.pdf}
  \BibitemShut {NoStop}%
\bibitem [{\citenamefont {Zhao}\ \emph {et~al.}(2008)\citenamefont {Zhao},
  \citenamefont {Marhefka}, \citenamefont {Shu}, \citenamefont {Hund},
  \citenamefont {Kameneva},\ and\ \citenamefont {Antaki}}]{Zhao2008}%
  \BibitemOpen
  \bibfield  {author} {\bibinfo {author} {\bibfnamefont {R.}~\bibnamefont
  {Zhao}}, \bibinfo {author} {\bibfnamefont {J.}~\bibnamefont {Marhefka}},
  \bibinfo {author} {\bibfnamefont {F.}~\bibnamefont {Shu}}, \bibinfo {author}
  {\bibfnamefont {S.}~\bibnamefont {Hund}}, \bibinfo {author} {\bibfnamefont
  {M.}~\bibnamefont {Kameneva}}, \ and\ \bibinfo {author} {\bibfnamefont
  {J.}~\bibnamefont {Antaki}},\ }\href@noop {} {\bibfield  {journal} {\bibinfo
  {journal} {Annals of Biomedical Engineering}\ }\textbf {\bibinfo {volume}
  {36}},\ \bibinfo {pages} {1130} (\bibinfo {year} {2008})}\BibitemShut
  {NoStop}%
\bibitem [{\citenamefont {Wang}\ \emph
  {et~al.}(2013{\natexlab{b}})\citenamefont {Wang}, \citenamefont {Diacovo},
  \citenamefont {Chen}, \citenamefont {Freund},\ and\ \citenamefont
  {King}}]{Wang_2013_thrombus}%
  \BibitemOpen
  \bibfield  {author} {\bibinfo {author} {\bibfnamefont {W.}~\bibnamefont
  {Wang}}, \bibinfo {author} {\bibfnamefont {T.~G.}\ \bibnamefont {Diacovo}},
  \bibinfo {author} {\bibfnamefont {J.}~\bibnamefont {Chen}}, \bibinfo {author}
  {\bibfnamefont {J.~B.}\ \bibnamefont {Freund}}, \ and\ \bibinfo {author}
  {\bibfnamefont {M.~R.}\ \bibnamefont {King}},\ }\href@noop {} {\bibfield
  {journal} {\bibinfo  {journal} {PLoS ONE}\ }\textbf {\bibinfo {volume} {8}},\
  \bibinfo {pages} {e76949} (\bibinfo {year} {2013}{\natexlab{b}})}\BibitemShut
  {NoStop}%
\bibitem [{\citenamefont {Skorczewski}\ \emph {et~al.}(2013)\citenamefont
  {Skorczewski}, \citenamefont {Erickson},\ and\ \citenamefont
  {Fogelson}}]{Skorczewski2013}%
  \BibitemOpen
  \bibfield  {author} {\bibinfo {author} {\bibfnamefont {T.}~\bibnamefont
  {Skorczewski}}, \bibinfo {author} {\bibfnamefont {L.}~\bibnamefont
  {Erickson}}, \ and\ \bibinfo {author} {\bibfnamefont {A.~L.}\ \bibnamefont
  {Fogelson}},\ }\href {\doibase http://dx.doi.org/10.1016/j.bpj.2013.01.061}
  {\bibfield  {journal} {\bibinfo  {journal} {Biophysical Journal}\ }\textbf
  {\bibinfo {volume} {104}},\ \bibinfo {pages} {1764 } (\bibinfo {year}
  {2013})}\BibitemShut {NoStop}%
\bibitem [{\citenamefont {Yazdani}\ and\ \citenamefont
  {Karniadakis}(2016)}]{Yazdani_2016}%
  \BibitemOpen
  \bibfield  {author} {\bibinfo {author} {\bibfnamefont {A.}~\bibnamefont
  {Yazdani}}\ and\ \bibinfo {author} {\bibfnamefont {G.~E.}\ \bibnamefont
  {Karniadakis}},\ }\href@noop {} {\bibfield  {journal} {\bibinfo  {journal}
  {Soft Matter}\ }\textbf {\bibinfo {volume} {12}},\ \bibinfo {pages} {4339}
  (\bibinfo {year} {2016})}\BibitemShut {NoStop}%
\bibitem [{\citenamefont {Limbach}\ \emph {et~al.}(2006)\citenamefont
  {Limbach}, \citenamefont {Arnold}, \citenamefont {Mann},\ and\ \citenamefont
  {Holm}}]{Limbach2006}%
  \BibitemOpen
  \bibfield  {author} {\bibinfo {author} {\bibfnamefont {H.}~\bibnamefont
  {Limbach}}, \bibinfo {author} {\bibfnamefont {A.}~\bibnamefont {Arnold}},
  \bibinfo {author} {\bibfnamefont {B.}~\bibnamefont {Mann}}, \ and\ \bibinfo
  {author} {\bibfnamefont {C.}~\bibnamefont {Holm}},\ }\href {\doibase
  http://dx.doi.org/10.1016/j.cpc.2005.10.005} {\bibfield  {journal} {\bibinfo
  {journal} {Computer Physics Communications}\ }\textbf {\bibinfo {volume}
  {174}},\ \bibinfo {pages} {704 } (\bibinfo {year} {2006})}\BibitemShut
  {NoStop}%
\bibitem [{\citenamefont {Roehm}\ and\ \citenamefont
  {Arnold}(2012)}]{Roehm2012}%
  \BibitemOpen
  \bibfield  {author} {\bibinfo {author} {\bibfnamefont {D.}~\bibnamefont
  {Roehm}}\ and\ \bibinfo {author} {\bibfnamefont {A.}~\bibnamefont {Arnold}},\
  }\href {\doibase 10.1140/epjst/e2012-01639-6} {\bibfield  {journal} {\bibinfo
   {journal} {The European Physical Journal Special Topics}\ }\textbf {\bibinfo
  {volume} {210}},\ \bibinfo {pages} {89} (\bibinfo {year} {2012})}\BibitemShut
  {NoStop}%
\bibitem [{\citenamefont {Arnold}\ \emph {et~al.}(2013)\citenamefont {Arnold},
  \citenamefont {Lenz}, \citenamefont {Kesselheim}, \citenamefont {Weeber},
  \citenamefont {Fahrenberger}, \citenamefont {Roehm}, \citenamefont
  {Ko{\v{s}}ovan},\ and\ \citenamefont {Holm}}]{Arnold2013}%
  \BibitemOpen
  \bibfield  {author} {\bibinfo {author} {\bibfnamefont {A.}~\bibnamefont
  {Arnold}}, \bibinfo {author} {\bibfnamefont {O.}~\bibnamefont {Lenz}},
  \bibinfo {author} {\bibfnamefont {S.}~\bibnamefont {Kesselheim}}, \bibinfo
  {author} {\bibfnamefont {R.}~\bibnamefont {Weeber}}, \bibinfo {author}
  {\bibfnamefont {F.}~\bibnamefont {Fahrenberger}}, \bibinfo {author}
  {\bibfnamefont {D.}~\bibnamefont {Roehm}}, \bibinfo {author} {\bibfnamefont
  {P.}~\bibnamefont {Ko{\v{s}}ovan}}, \ and\ \bibinfo {author} {\bibfnamefont
  {C.}~\bibnamefont {Holm}},\ }in\ \href@noop {} {\emph {\bibinfo {booktitle}
  {Meshfree methods for partial differential equations VI}}}\ (\bibinfo
  {publisher} {Springer Berlin Heidelberg},\ \bibinfo {year} {2013})\ pp.\
  \bibinfo {pages} {1--23}\BibitemShut {NoStop}%
\bibitem [{\citenamefont {Peskin}(2002)}]{Peskin2002}%
  \BibitemOpen
  \bibfield  {author} {\bibinfo {author} {\bibfnamefont {C.~S.}\ \bibnamefont
  {Peskin}},\ }\href@noop {} {\bibfield  {journal} {\bibinfo  {journal} {Acta
  numerica}\ }\textbf {\bibinfo {volume} {11}},\ \bibinfo {pages} {479}
  (\bibinfo {year} {2002})}\BibitemShut {NoStop}%
\bibitem [{\citenamefont {Mittal}\ and\ \citenamefont
  {Iaccarino}(2005)}]{Mittal2005}%
  \BibitemOpen
  \bibfield  {author} {\bibinfo {author} {\bibfnamefont {R.}~\bibnamefont
  {Mittal}}\ and\ \bibinfo {author} {\bibfnamefont {G.}~\bibnamefont
  {Iaccarino}},\ }\href {\doibase 10.1146/annurev.fluid.37.061903.175743}
  {\bibfield  {journal} {\bibinfo  {journal} {Annual Review of Fluid
  Mechanics}\ }\textbf {\bibinfo {volume} {37}},\ \bibinfo {pages} {239}
  (\bibinfo {year} {2005})},\ \Eprint
  {http://arxiv.org/abs/http://dx.doi.org/10.1146/annurev.fluid.37.061903.175743}
  {http://dx.doi.org/10.1146/annurev.fluid.37.061903.175743} \BibitemShut
  {NoStop}%
\bibitem [{\citenamefont {Succi}(2001)}]{Succi2001}%
  \BibitemOpen
  \bibfield  {author} {\bibinfo {author} {\bibfnamefont {S.}~\bibnamefont
  {Succi}},\ }\href@noop {} {\emph {\bibinfo {title} {The lattice Boltzmann
  equation: for fluid dynamics and beyond}}}\ (\bibinfo  {publisher} {Oxford
  university press},\ \bibinfo {year} {2001})\BibitemShut {NoStop}%
\bibitem [{\citenamefont {D{\"u}nweg}\ and\ \citenamefont
  {Ladd}(2008)}]{Duenweg2008}%
  \BibitemOpen
  \bibfield  {author} {\bibinfo {author} {\bibfnamefont {B.}~\bibnamefont
  {D{\"u}nweg}}\ and\ \bibinfo {author} {\bibfnamefont {A.~J.}\ \bibnamefont
  {Ladd}},\ }\enquote {\bibinfo {title} {Advanced computer simulation
  approaches for soft matter sciences iii},}\ \ (\bibinfo  {publisher}
  {Springer},\ \bibinfo {year} {2008})\ Chap.\ \bibinfo {chapter} {Lattice
  Boltzmann Simulations of Soft Matter Systems}, pp.\ \bibinfo {pages}
  {89--166}\BibitemShut {NoStop}%
\bibitem [{\citenamefont {Aidun}\ and\ \citenamefont
  {Clausen}(2010)}]{Aidun2010}%
  \BibitemOpen
  \bibfield  {author} {\bibinfo {author} {\bibfnamefont {C.~K.}\ \bibnamefont
  {Aidun}}\ and\ \bibinfo {author} {\bibfnamefont {J.~R.}\ \bibnamefont
  {Clausen}},\ }\href {\doibase 10.1146/annurev-fluid-121108-145519} {\bibfield
   {journal} {\bibinfo  {journal} {Annual Review of Fluid Mechanics}\ }\textbf
  {\bibinfo {volume} {42}},\ \bibinfo {pages} {439} (\bibinfo {year} {2010})},\
  \Eprint
  {http://arxiv.org/abs/http://dx.doi.org/10.1146/annurev-fluid-121108-145519}
  {http://dx.doi.org/10.1146/annurev-fluid-121108-145519} \BibitemShut
  {NoStop}%
\bibitem [{\citenamefont {Barthès-Biesel}(2011)}]{BarthesBiesel2011}%
  \BibitemOpen
  \bibfield  {author} {\bibinfo {author} {\bibfnamefont {D.}~\bibnamefont
  {Barthès-Biesel}},\ }\href {\doibase
  http://dx.doi.org/10.1016/j.cocis.2010.07.001} {\bibfield  {journal}
  {\bibinfo  {journal} {Current Opinion in Colloid \& Interface Science}\
  }\textbf {\bibinfo {volume} {16}},\ \bibinfo {pages} {3 } (\bibinfo {year}
  {2011})}\BibitemShut {NoStop}%
\bibitem [{\citenamefont {Freund}(2013)}]{Freund2013}%
  \BibitemOpen
  \bibfield  {author} {\bibinfo {author} {\bibfnamefont {J.~B.}\ \bibnamefont
  {Freund}},\ }\href {\doibase http://dx.doi.org/10.1063/1.4819341} {\bibfield
  {journal} {\bibinfo  {journal} {Physics of Fluids}\ }\textbf {\bibinfo
  {volume} {25}},\ \bibinfo {eid} {110807} (\bibinfo {year} {2013}),\
  http://dx.doi.org/10.1063/1.4819341}\BibitemShut {NoStop}%
\bibitem [{\citenamefont {Farutin}\ \emph {et~al.}(2014)\citenamefont
  {Farutin}, \citenamefont {Biben},\ and\ \citenamefont
  {Misbah}}]{Farutin2014}%
  \BibitemOpen
  \bibfield  {author} {\bibinfo {author} {\bibfnamefont {A.}~\bibnamefont
  {Farutin}}, \bibinfo {author} {\bibfnamefont {T.}~\bibnamefont {Biben}}, \
  and\ \bibinfo {author} {\bibfnamefont {C.}~\bibnamefont {Misbah}},\ }\href
  {\doibase http://dx.doi.org/10.1016/j.jcp.2014.07.008} {\bibfield  {journal}
  {\bibinfo  {journal} {Journal of Computational Physics}\ }\textbf {\bibinfo
  {volume} {275}},\ \bibinfo {pages} {539 } (\bibinfo {year}
  {2014})}\BibitemShut {NoStop}%
\bibitem [{\citenamefont {Guckenberger}\ \emph {et~al.}(2016)\citenamefont
  {Guckenberger}, \citenamefont {Schraml}, \citenamefont {Chen}, \citenamefont
  {Leonetti},\ and\ \citenamefont {Gekle}}]{Guckenberger2016}%
  \BibitemOpen
  \bibfield  {author} {\bibinfo {author} {\bibfnamefont {A.}~\bibnamefont
  {Guckenberger}}, \bibinfo {author} {\bibfnamefont {M.~P.}\ \bibnamefont
  {Schraml}}, \bibinfo {author} {\bibfnamefont {P.~G.}\ \bibnamefont {Chen}},
  \bibinfo {author} {\bibfnamefont {M.}~\bibnamefont {Leonetti}}, \ and\
  \bibinfo {author} {\bibfnamefont {S.}~\bibnamefont {Gekle}},\ }\href
  {\doibase http://dx.doi.org/10.1016/j.cpc.2016.04.018} {\bibfield  {journal}
  {\bibinfo  {journal} {Computer Physics Communications}\ ,\ } (\bibinfo {year}
  {2016})}\BibitemShut {NoStop}%
\bibitem [{\citenamefont {Coupier}\ \emph {et~al.}(2008)\citenamefont
  {Coupier}, \citenamefont {Kaoui}, \citenamefont {Podgorski},\ and\
  \citenamefont {Misbah}}]{Coupier_2008}%
  \BibitemOpen
  \bibfield  {author} {\bibinfo {author} {\bibfnamefont {G.}~\bibnamefont
  {Coupier}}, \bibinfo {author} {\bibfnamefont {B.}~\bibnamefont {Kaoui}},
  \bibinfo {author} {\bibfnamefont {T.}~\bibnamefont {Podgorski}}, \ and\
  \bibinfo {author} {\bibfnamefont {C.}~\bibnamefont {Misbah}},\ }\href@noop {}
  {\bibfield  {journal} {\bibinfo  {journal} {Phys. Fluids}\ }\textbf {\bibinfo
  {volume} {20}},\ \bibinfo {pages} {111702} (\bibinfo {year}
  {2008})}\BibitemShut {NoStop}%
\bibitem [{\citenamefont {Geislinger}\ \emph {et~al.}(2012)\citenamefont
  {Geislinger}, \citenamefont {Eggart}, \citenamefont {Braunm{\"u}ller},
  \citenamefont {Schmid},\ and\ \citenamefont {Franke}}]{Geislinger_2012}%
  \BibitemOpen
  \bibfield  {author} {\bibinfo {author} {\bibfnamefont {T.~M.}\ \bibnamefont
  {Geislinger}}, \bibinfo {author} {\bibfnamefont {B.}~\bibnamefont {Eggart}},
  \bibinfo {author} {\bibfnamefont {S.}~\bibnamefont {Braunm{\"u}ller}},
  \bibinfo {author} {\bibfnamefont {L.}~\bibnamefont {Schmid}}, \ and\ \bibinfo
  {author} {\bibfnamefont {T.}~\bibnamefont {Franke}},\ }\href@noop {}
  {\bibfield  {journal} {\bibinfo  {journal} {Appl. Phys. Lett.}\ }\textbf
  {\bibinfo {volume} {100}},\ \bibinfo {pages} {183701} (\bibinfo {year}
  {2012})}\BibitemShut {NoStop}%
\bibitem [{\citenamefont {Grandchamp}\ \emph {et~al.}(2013)\citenamefont
  {Grandchamp}, \citenamefont {Coupier}, \citenamefont {Srivastav},
  \citenamefont {Minetti},\ and\ \citenamefont {Podgorski}}]{Grandchamp_2013}%
  \BibitemOpen
  \bibfield  {author} {\bibinfo {author} {\bibfnamefont {X.}~\bibnamefont
  {Grandchamp}}, \bibinfo {author} {\bibfnamefont {G.}~\bibnamefont {Coupier}},
  \bibinfo {author} {\bibfnamefont {A.}~\bibnamefont {Srivastav}}, \bibinfo
  {author} {\bibfnamefont {C.}~\bibnamefont {Minetti}}, \ and\ \bibinfo
  {author} {\bibfnamefont {T.}~\bibnamefont {Podgorski}},\ }\href@noop {}
  {\bibfield  {journal} {\bibinfo  {journal} {Phys. Rev. Lett.}\ }\textbf
  {\bibinfo {volume} {110}},\ \bibinfo {pages} {108101} (\bibinfo {year}
  {2013})}\BibitemShut {NoStop}%
\bibitem [{\citenamefont {Fedosov}\ \emph {et~al.}(2010)\citenamefont
  {Fedosov}, \citenamefont {Caswell}, \citenamefont {Popel},\ and\
  \citenamefont {Karniadakis}}]{Fedosov_2010_CFL}%
  \BibitemOpen
  \bibfield  {author} {\bibinfo {author} {\bibfnamefont {D.~A.}\ \bibnamefont
  {Fedosov}}, \bibinfo {author} {\bibfnamefont {B.}~\bibnamefont {Caswell}},
  \bibinfo {author} {\bibfnamefont {A.~S.}\ \bibnamefont {Popel}}, \ and\
  \bibinfo {author} {\bibfnamefont {G.~E.}\ \bibnamefont {Karniadakis}},\
  }\href@noop {} {\bibfield  {journal} {\bibinfo  {journal} {Microcirculation}\
  }\textbf {\bibinfo {volume} {17}},\ \bibinfo {pages} {615} (\bibinfo {year}
  {2010})}\BibitemShut {NoStop}%
\bibitem [{\citenamefont {Freund}\ and\ \citenamefont
  {Orescanin}(2011)}]{Freund_2011}%
  \BibitemOpen
  \bibfield  {author} {\bibinfo {author} {\bibfnamefont {J.~B.}\ \bibnamefont
  {Freund}}\ and\ \bibinfo {author} {\bibfnamefont {M.~M.}\ \bibnamefont
  {Orescanin}},\ }\href@noop {} {\bibfield  {journal} {\bibinfo  {journal} {J
  Fluid Mech}\ }\textbf {\bibinfo {volume} {671}},\ \bibinfo {pages} {466}
  (\bibinfo {year} {2011})}\BibitemShut {NoStop}%
\bibitem [{\citenamefont {Katanov}\ \emph {et~al.}(2015)\citenamefont
  {Katanov}, \citenamefont {Gompper},\ and\ \citenamefont
  {Fedosov}}]{Katanov_2015}%
  \BibitemOpen
  \bibfield  {author} {\bibinfo {author} {\bibfnamefont {D.}~\bibnamefont
  {Katanov}}, \bibinfo {author} {\bibfnamefont {G.}~\bibnamefont {Gompper}}, \
  and\ \bibinfo {author} {\bibfnamefont {D.~A.}\ \bibnamefont {Fedosov}},\
  }\href@noop {} {\bibfield  {journal} {\bibinfo  {journal} {Microvasc. Res.}\
  }\textbf {\bibinfo {volume} {99}},\ \bibinfo {pages} {57} (\bibinfo {year}
  {2015})}\BibitemShut {NoStop}%
\bibitem [{\citenamefont {Fogelson}\ and\ \citenamefont
  {Neeves}(2015)}]{Fogelson_2015}%
  \BibitemOpen
  \bibfield  {author} {\bibinfo {author} {\bibfnamefont {A.~L.}\ \bibnamefont
  {Fogelson}}\ and\ \bibinfo {author} {\bibfnamefont {K.~B.}\ \bibnamefont
  {Neeves}},\ }\href@noop {} {\bibfield  {journal} {\bibinfo  {journal} {Annu.
  Rev. Fluid Mech.}\ }\textbf {\bibinfo {volume} {47}},\ \bibinfo {pages} {377}
  (\bibinfo {year} {2015})}\BibitemShut {NoStop}%
\end{thebibliography}%
